\documentstyle[psfig]{elsart}
\def\la{\mathrel{\hbox{\rlap{\hbox{\lower4pt\hbox{$\sim$}}}\hbox{$<$}}}}
\def\ga{\mathrel{\hbox{\rlap{\hbox{\lower4pt\hbox{$\sim$}}}\hbox{$>$}}}}

\def\be{\begin{equation}}
\def\ee{\end{equation}}
\def\bea{\begin{eqnarray}}
\def\eea{\end{eqnarray}}

\def\sec{\ifmmode \,\, {\rm sec} \else sec \fi}
\def\eV {\ifmmode \,\, {\rm eV} \else eV \fi}
\def\keV{\ifmmode \,\, {\rm keV} \else keV \fi}
\def\MeV{\ifmmode \,\, {\rm MeV} \else MeV \fi}
\def\GeV{\ifmmode \,\, {\rm GeV} \else GeV \fi}
\def\TeV{\ifmmode \,\, {\rm TeV} \else TeV \fi}
\def\fm{\ifmmode \,\, {\rm fm} \else TeV \fi}
\def\pbarn{\ifmmode \,\, {\rm pb} \else pb \fi}
\def\km{\ifmmode {\rm km}\, \else km \fi}
\def\Mpc{\ifmmode {\rm Mpc}\, \else Mpc \fi}
\def\Gyr{\ifmmode {\rm Gyr}\, \else Gyr \fi}

\def\fun#1#2{\lower3.6pt\vbox{\baselineskip0pt\lineskip.9pt
  \ialign{$\mathsurround=0pt#1\hfil##\hfil$\crcr#2\crcr\sim\crcr}}}
\def\la{\mathrel{\mathpalette\fun <}}
\def\ga{\mathrel{\mathpalette\fun >}}

\def\sbar#1{\kern 0.8pt
        \overline{\kern -0.8pt #1 \kern -0.8pt}
        \kern 0.8pt}  

\def\meter{\ifmmode \,\, {\rm m} \else m \fi}
\def\yr {\ifmmode \,\, {\rm yr} \else yr \fi}

\def\sr{\ifmmode \,\, {\rm sr} \else sr \fi}

\def\slashchar#1{\setbox0=\hbox{$#1$}           
   \dimen0=\wd0                                 
   \setbox1=\hbox{/} \dimen1=\wd1               
   \ifdim\dimen0>\dimen1                        
      \rlap{\hbox to \dimen0{\hfil/\hfil}}      
      #1                                        
   \else					
      \rlap{\hbox to \dimen1{\hfil$#1$\hfil}}   
      /                                         
   \fi}

\begin{document}
\begin{frontmatter}
\title{Implications of Recent Observational Discoveries for the Nature
and Origin of Gamma-Ray Bursts}
\author{D.~Q.~Lamb}
\address{Department of Astronomy \& Astrophysics, University of
Chicago, Chicago, IL 60637, U.S.A.}
\thanks[dqlemail]{E-mail: lamb@oddjob.uchicago.edu}

\begin{abstract} 
The discoveries that GRBs have X-ray, optical and radio afterglows have
connected the study of GRBs to the rest of astronomy, and
revolutionized the field.  In this review, I discuss the implications 
that the observation of these afterglows have for burst energies and
luminosities, and for models of the bursts and their afterglows.  I
describe recent evidence linking the long, softer, smoother GRBs
detected by BeppoSAX  and core collapse supernovae.  Finally, I
summarize recent work showing that, if these GRBs are due to the
collapse of massive stars, they may provide a powerful probe of the
very high redshift universe.
\end{abstract}
\begin{keyword}
Gamma rays: bursts
\end{keyword}
\end{frontmatter} 

\section{Introduction}

Gamma-ray bursts (GRBs) were discovered serendipitously more than a
quarter century ago~\cite{kleb73}.  The bursts consist of short,
intense episodes of gamma-ray emission, lasting anywhere from $\sim
10^{-2}$ seconds to $\sim 10^3$ seconds.  The time histories of GRBs
are diverse, as Figure 1 illustrates.  However, studies of the time
histories have shown that the bursts can be separated into two classes:
short, harder, more variable bursts; and long, softer, smoother 
bursts~\cite{lamb93,kouveliotou93} (see Figure 2).  Burst spectra are
nonthermal; the photon number spectrum is a broken power law, with 
average slopes $\sim -1.5$ and $\sim -2.5$ at low and high energies and
a shoulder at 100 keV - 1 Mev~\cite{band93,schaefer94}.

The data gathered by the Burst and Transient Source Experiment (BATSE)
on the {\it Compton} Gamma-Ray Observatory confirmed earlier evidence
of a rollover in the cumulative brightness distribution of GRBs,
showing that the burst sources are inhomogeneously distributed in
space~\cite{meegan92}.  The data also showed that the sky distribution
of even faint bursts is consistent with isotropy~\cite{meegan92} (see
Figure 3).  This combination of results implies that we are at, or
near, the center of the spatial distribution of burst sources and that
the intrinsic brightness and/or spatial density of the sources
decreases with increasing distance from us.

\begin{figure}[t]
\hfil
\begin{minipage}[t]{2.5truein}
\mbox{}\\
\psfig{file=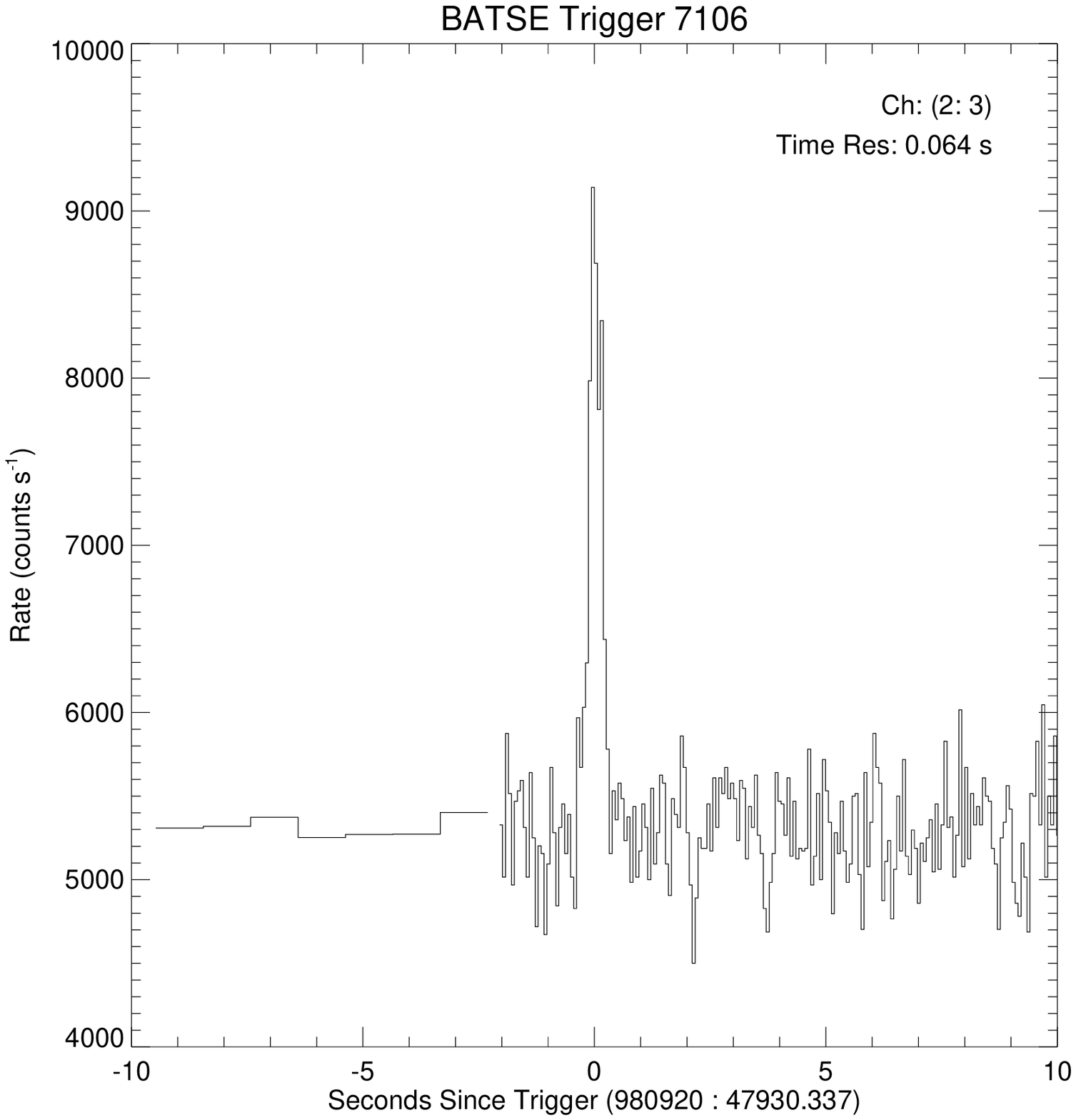,width=2.5truein,clip=}
\end{minipage}
\hfill
\begin{minipage}[t]{2.5truein}
\mbox{}\\
\hfil
\psfig{file=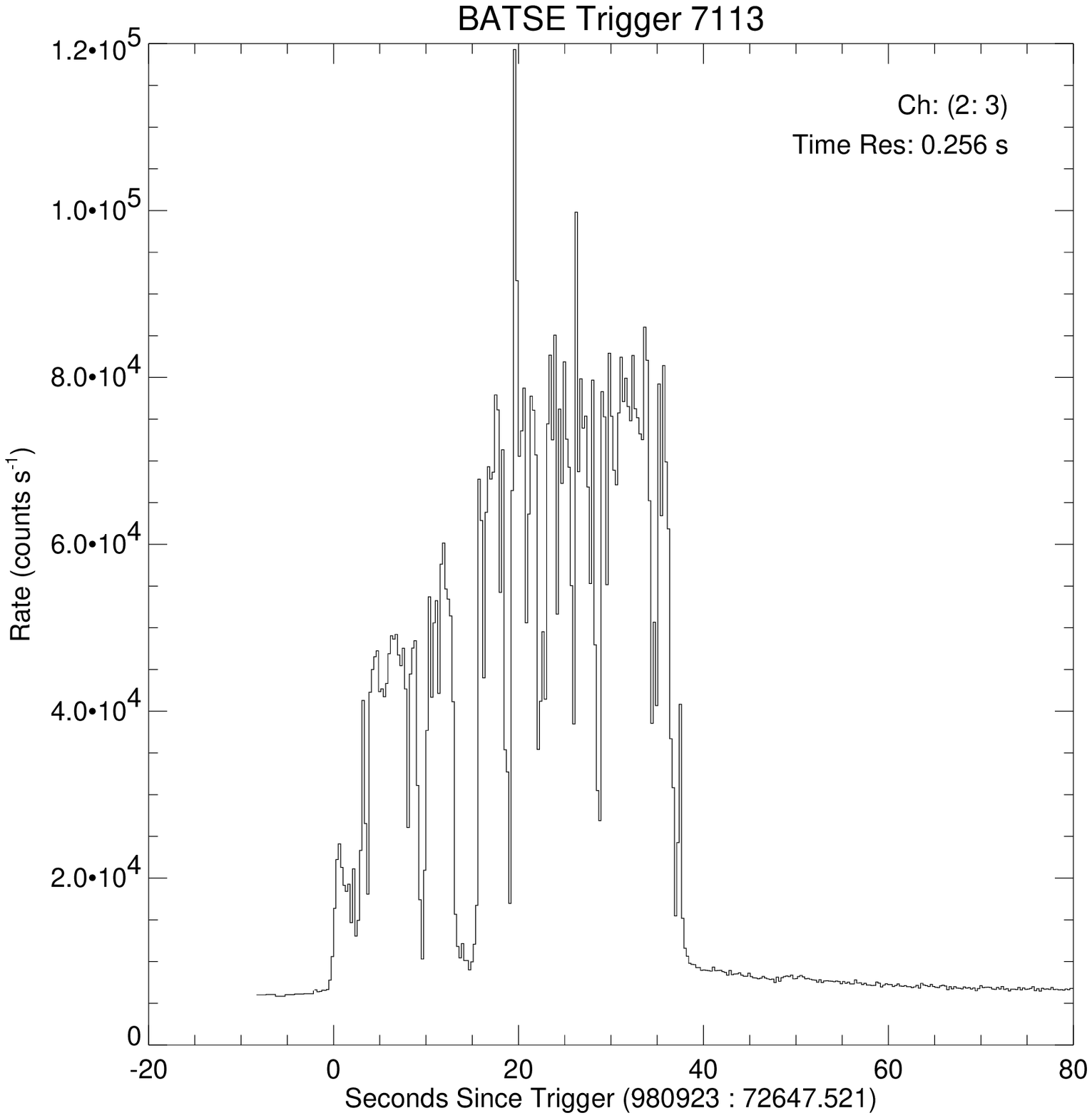,width=2.5truein,clip=}
\hfil
\vfil
\end{minipage}
\\
\begin{minipage}[t]{2.5truein}
\mbox{}\\
\psfig{file=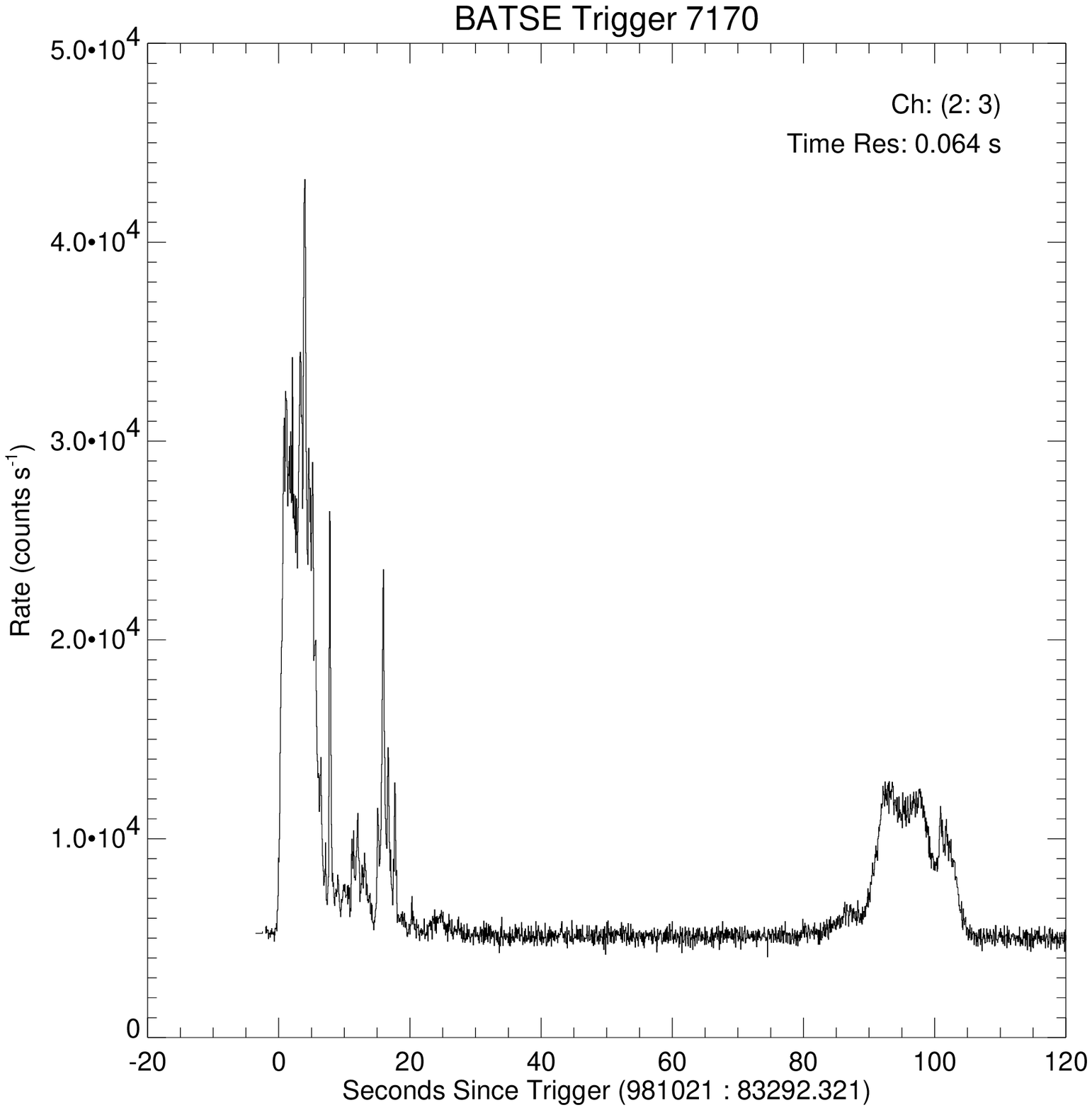,width=2.5truein,clip=}
\end{minipage}
\hfill
\begin{minipage}[t]{2.5truein}
\mbox{}\\
\psfig{file=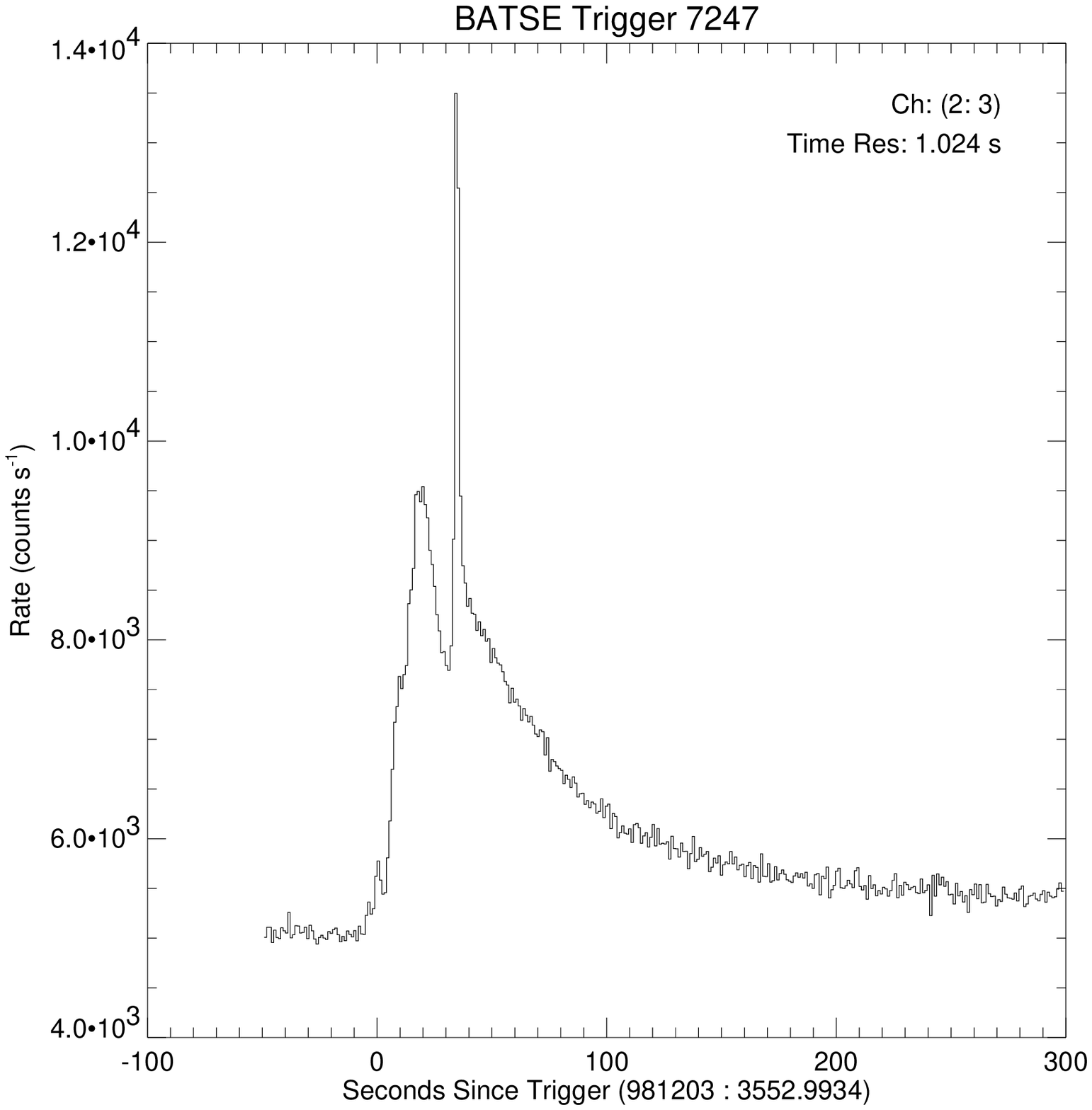,width=2.5truein,clip=}
\vfil
\end{minipage}
\caption{Four GRB time histories of GRBs 980920, 980923, 981021, and
981203.  These four bursts occurred within about two months,
illustrating the diversity of GRB time histories.~\cite{paciesas99}}
\end{figure}

\begin{figure}[t]
\hfil
\psfig{file=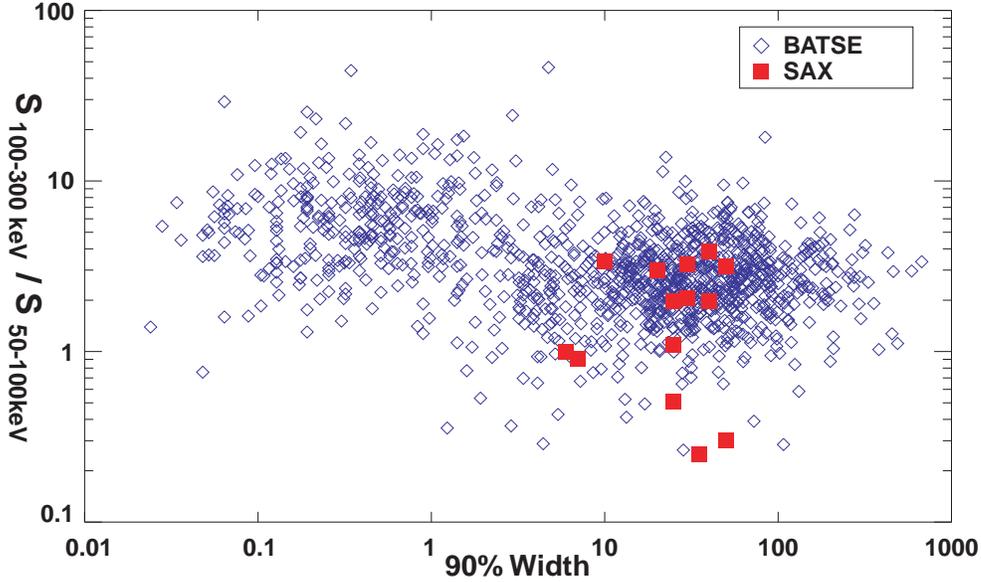,width=5.12truein,clip=}
\hfil
\caption{Distribution of duration (as measured by the time interval
containing 90\% of the photon counts) versus spectral hardness (as
measured by the ratio of fluence in the 50 - 100 keV and 100 - 300 keV
energy bands) for bursts in the BATSE 4B catalog~\cite{paciesas99} 
(diamonds), showing clear evidence for two classes of bursts: short,
harder, more variable bursts; and long, softer, smoother bursts. 
Events detected by BeppoSAX (solid squares) belong to the latter
class.  From~\cite{kulkarni00}.}
\end{figure}

\begin{figure}[b]
\hfil
\psfig{file=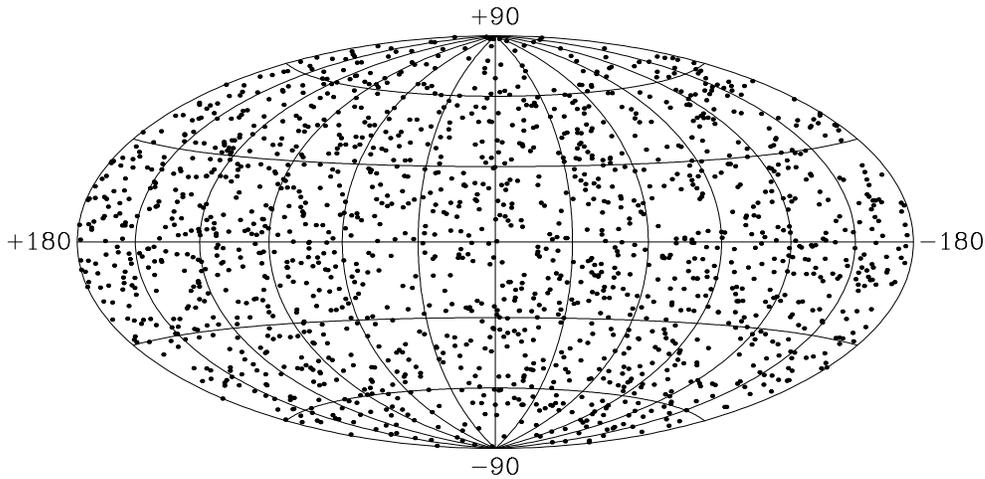,width=5.12truein,clip=}
\hfil
\caption{The positions in Galactic coordinates of the GRBs in the BATSE
4B catalog~\cite{paciesas99}, showing the isotropy of the burst sky
distribution~\cite{meegan92}.}
\end{figure}

The BATSE results showed that the bursts cannot come from a population
of neutron stars in a thick Galactic disk (as was previously thought) 
and spurred interest in the possibility that the sources of the bursts
lie at cosmological distances.  Yet the evidence remained
circumstantial.  Consequently, the GRB distance scale -- and even more,
the nature of the burst sources -- was debatable~\cite{lamb95,pac95}.  
The principal reason for the continuing uncertainty in the distance
scale of the bursts was that no definite counterpart to any burst
could be found at other wavelengths, despite intense efforts spanning
more than two decades.  Consequently, the study of GRBs was isolated
from the rest of astronomy.  Scientists studying them had only the laws
of physics and the properties of the bursts themselves to guide them in
attempting to solve the GRB mystery.

\begin{figure}[t]
\hfil
\psfig{file=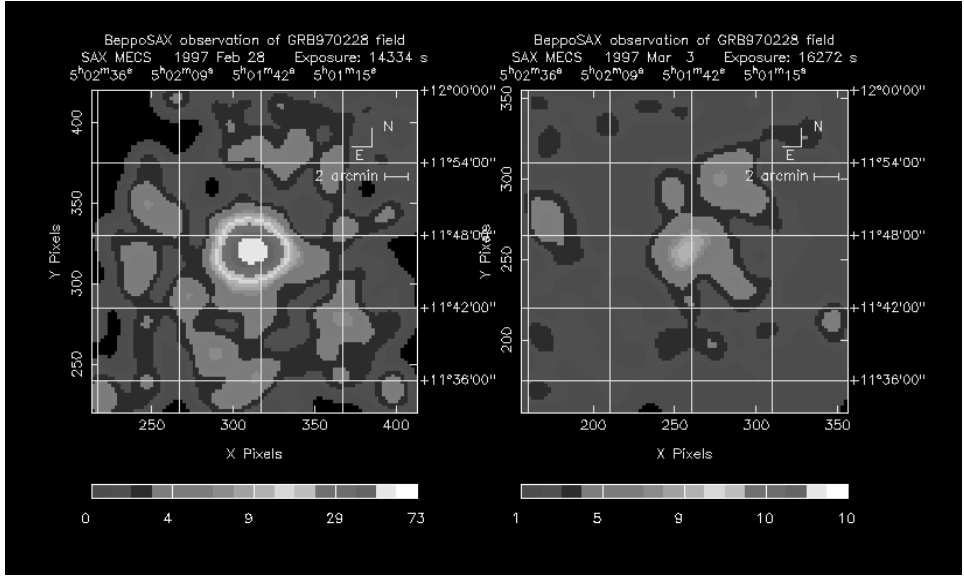,width=5truein,clip=}
\caption{BeppoSAX observations of the fading X-ray afterglow of GRB
970228. Left panel: MECS image on February 28.  Right panel: a deeper
MECS image on March 3~\cite{costa97b}}
\hfil
\end{figure}

\begin{figure}[b]
\hfil
\psfig{file=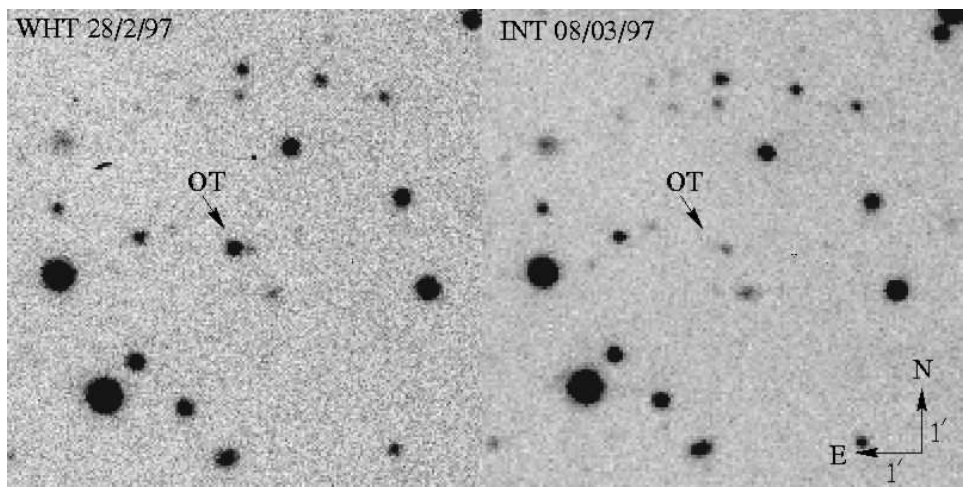,angle=270,width=5truein,clip=}
\hfil
\caption{William Herschel Telescope images taken on February 28 and
March 8 of the fading optical afterglow of GRB 970228.~\cite{paradijs97}}
\end{figure}

\begin{figure}[t]
\hfill
\begin{minipage}[t]{2.16truein}
\mbox{}\\
\psfig{file=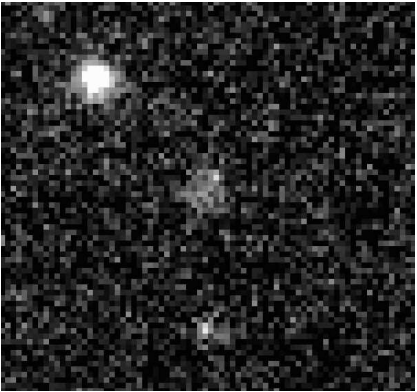,width=2.16truein,clip=}
\end{minipage}
\hfill
\begin{minipage}[t]{2.66truein}
\mbox{}\\
\caption{HST STIS image taken on September 4 of the optical afterglow of
GRB 970228, which revealed the presence of a faint ($R = 25.5$) host
galaxy.  Both are clearly visible near the center of the WFPC2 
image~\cite{sahu97}.}
\end{minipage}
\end{figure}

The relatively accurate (3') gamma-ray burst positions found using
BeppoSAX, and disseminated within a day or so, revolutionized the
field.  They led to the remarkable discoveries that GRBs have
X-ray~\cite{costa97}, optical~\cite{paradijs97} and radio~\cite{frail97}
afterglows, finally connecting the study of GRBs with the rest of
astronomy.  The breakthroughs in our understanding of GRBs made
possible by these discoveries cannot be overstated. 

In this review, dedicated to the memory of my friend and colleague
David Schramm, I describe these recent breakthroughs.  I first relate
the discovery of GRB X-ray, optical and radio afterglows.  I then
discuss the implications that the observations of these afterglows have
for burst energies and luminosities, and for models of the bursts and
their afterglows.  I describe recent evidence linking GRBs and core
collapse supernovae.  Finally, I summarize recent work showing that, if
GRBs are due to the collapse of massive stars, GRBs may provide a
powerful probe of the very high redshift universe.

\section{Recent Discoveries}

\begin{figure}[b]
\begin{minipage}[b]{2.36truein}
\mbox{}\\
\psfig{file=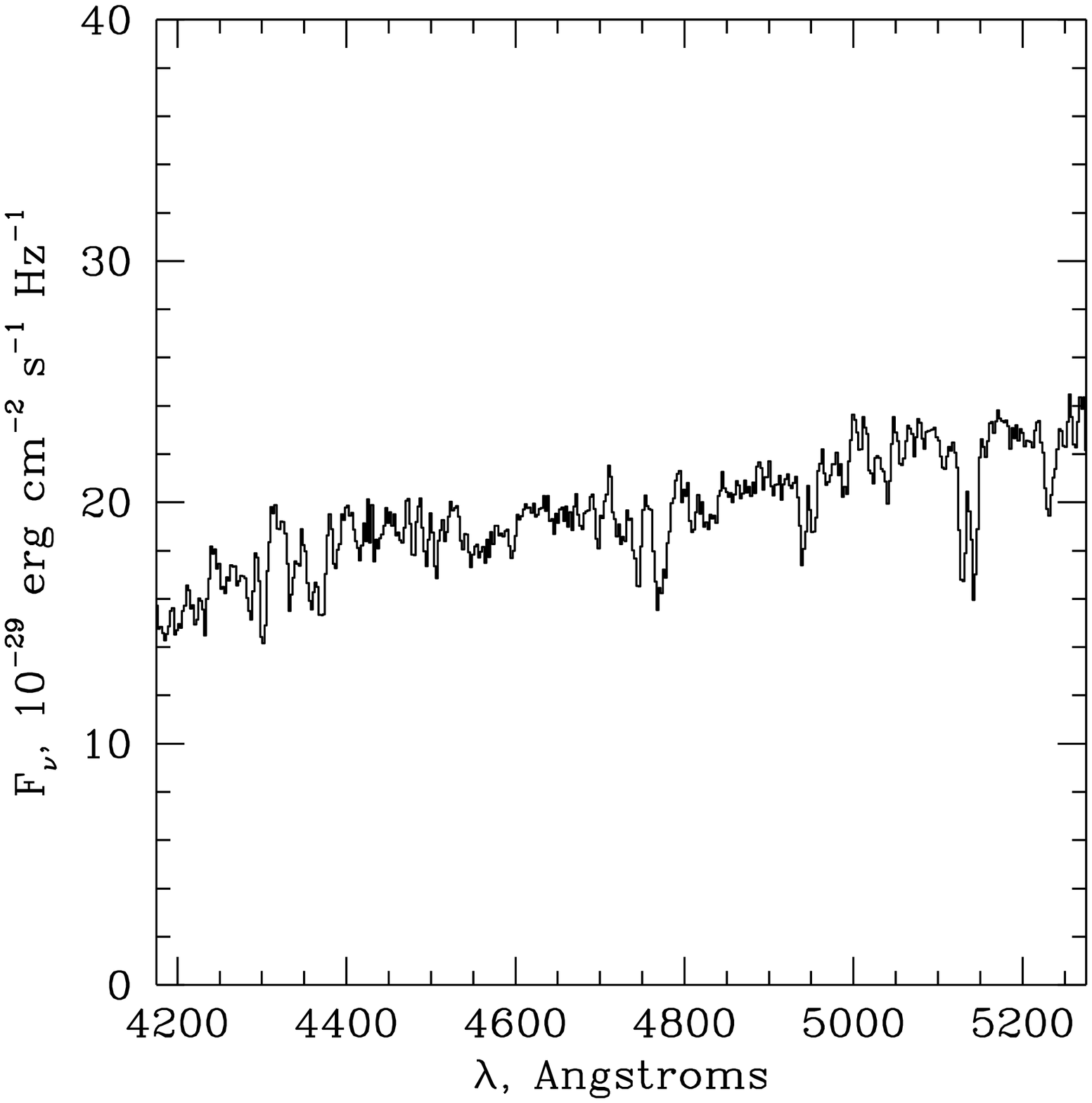,width=2.36truein,clip=}
\caption{Absorption lines of [Mg II] 5130 and 5144 \AA\ and [Fe II] 4770
\AA, among others, at a redshift $z = 0.83$ in the spectrum of the
optical afterglow of GRB 970508, showing that the burst occurred at
this redshift or larger.~\cite{metzger97}}
\end{minipage}
\hfill
\begin{minipage}[b]{2.76truein}
\mbox{}\\
\psfig{file=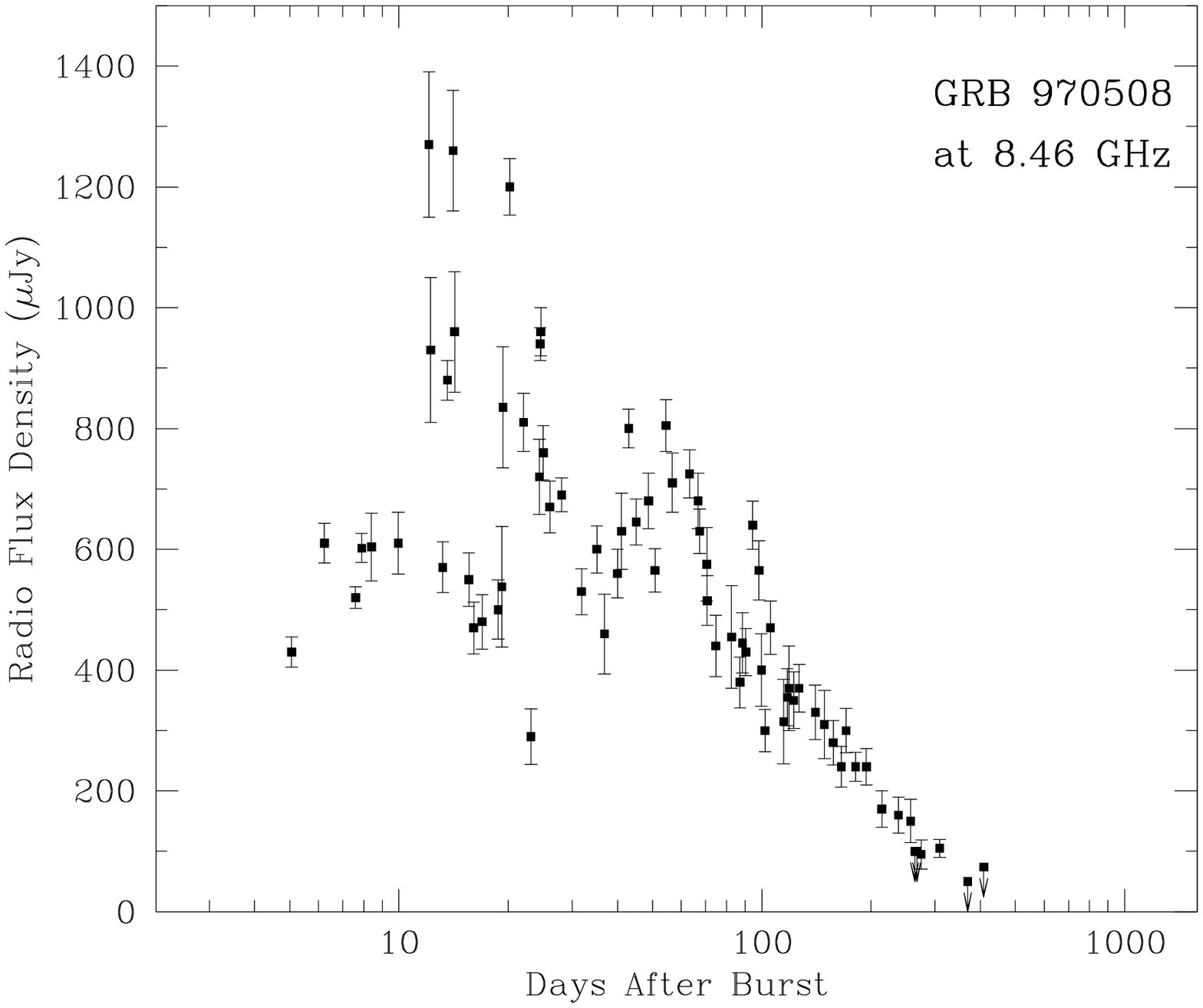,width=2.76truein,clip=} \caption{Radio
observations of the afterglow of GRB 970508.  The rapid variability
seen at early times disappears after $\approx$ 30 days.  This behavior
has been interpreted as due to scintillation, implying that GRB sources
are highly compact and expand at relativistic speeds~\cite{frail97}.}
\end{minipage}
\end{figure}

Following the detection of GRB 970228 by BeppoSAX, follow-up
observations revealed a fading X-ray source coincident with the
position of the GRB~\cite{costa97} (see Figure 4).  The relatively
rapid dissemination of this position led to the discovery of a fading
optical source coincident with the X-ray source~\cite{paradijs97} (see
Figure 5), establishing that GRBs exhibit both X-ray and optical
afterglows.  Subsequent observations using the Hubble Space Telescope
revealed a faint (R = 25.5) underlying galaxy~\cite{sahu97} (see
Figure 6).

\begin{figure}
\begin{minipage}[t]{2.76truein}
\mbox{}\\
\psfig{file=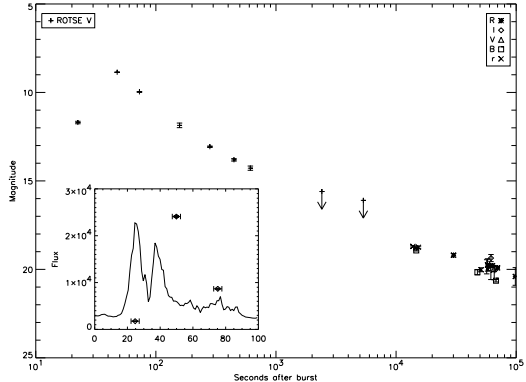,angle=90,width=2.76truein,clip=}
\end{minipage}
\hfill
\begin{minipage}[t]{2.36truein}
\mbox{}\\
\caption{ROTSE observations of bright ($m_V \approx 9$) optical
emission coincident with GRB 990123.  The inset shows the BATSE time
history of the burst and three of the ROTSE data points (arbitrarily
scaled to the BATSE time history)~\cite{akerlof99}.}
\end{minipage}
\end{figure}

Two months later, Metzger et al.~\cite{metzger97} found redshifted
absorption lines of Mg II and Fe II at $z = 0.83$ in the optical
spectrum of the GRB 970508 afterglow (see Figure 7), which established
that most (perhaps all) GRB sources lie at cosmological distances.  At
about the same time, Frail et al.~\cite{frail97} discovered a radio
source coincident with the fading X-ray afterglow of this burst,
marking the first detection of a radio afterglow.  The radio afterglow
showed rapid variability, which disappeared after about 30 days (see
Figure 8).  The rapid variability has been interpreted as due to
scintillation, implying that GRB sources are highly compact and expand
at relativistic speeds~\cite{frail97}.

Finally, we now know that GRBs can produce astonishingly bright optical
emission coincident with the burst itself.  ROTSE, a robotic optical
telescope, slewed to the position of GRB 990123, a burst whose position
was determined by BATSE and disseminated by the Gamma-Ray Burst
Coordinate Network (GCN), within $\approx 20$ seconds of the onset of
the burst.  The ROTSE observations revealed intense optical emission,
peaking at $m_V \approx 9$ during the burst and fading
thereafter~\cite{akerlof99} (see Figure 9).

X-ray afterglows are now known for about two dozen GRBs~\cite{costa99}.
Optical afterglows have been detected for roughly half of these
bursts~\cite{kulkarni00}, and radio afterglows have been detected for
about one third of them~\cite{frail00}.  Host galaxies have been found
for most of the bursts with detected optical and/or radio
afterglows~\cite{kulkarni00}.  Redshifts are known for eight GRBs,
three of these are from spectroscopic observations of the optical
afterglow, and the remainder are from spectroscopic observations of the
host galaxy (see Table 1).

Not so long ago, adherents of the cosmological hypothesis for GRBs
favored a redshift range $0.1 \la z \la 1$, which was derived primarily
from the brightness distribution of the bursts under the assumption
that GRBs are standard candles (see, e.g.,~\cite{fenimore93}).  Now
astrophysicists routinely talk about redshift distances $1 \la z \la
6$, a redshift range that is consistent with the modest number of GRB
redshifts that have been determined so far (see,
e.g,~\cite{totani97,totani99,wijers99,hogg99}).  The increase in the
GRB distance scale implies that the GRB phenomenon is much rarer than
was previously thought.  For example, Schmidt~\cite{schmidt00} finds
that the GRB rate must be
\begin{equation}
R_{\rm GRB} \sim 10^{-10}\ {\rm GRBs}\ {\rm yr}^{-1}\ {\rm Mpc}^{-3}
\end{equation}
in order both to match the brightness distribution of the BATSE bursts
and to accommodate the redshift distance of $z = 3.42$ inferred for GRB
971214.

By comparison, the rates of neutron star-neutron star (NS-NS) binary
mergers~\cite{totani99} and the rate of Type Ib-Ic
supernovae~\cite{cap97} are
\begin{eqnarray}
R_{\rm NS-NS} \sim 10^{-6}\ {\rm mergers}\ {\rm yr}^{-1}\ {\rm Mpc}^{-3}
\\ \nonumber
\\
R_{\rm Type\ Ib-Ic} \sim 3 \times 10^{-5}\ {\rm SNe}\ {\rm yr}^{-1}\ 
{\rm Mpc}^{-3}  \; .
\end{eqnarray}
The rate of neutron star-black hole (NS-BH) binary mergers will be
smaller.  Nevertheless, it is clear that, if either of these events are
the sources of GRBs, only a tiny fraction of them produce an observable
GRB.  Even if one posits strong collimation (i.e., $f_{\rm jet} \approx
10^{-2}$), the fraction is small:
\begin{eqnarray}
R_{\rm GRB}/R_{\rm NS-NS} \sim 10^{-2}\ (f_{\rm jet}/10^{-2})^{-1} 
\\ \nonumber
\\
R_{\rm GRB}/R_{\rm Type\ Ib-Ic} \sim 3 \times 10^{-4}\ 
(f_{\rm jet}/10^{-2})^{-1} \; .
\end{eqnarray}
Therefore, if such events are the sources of GRBs, either the bursts
are very strongly collimated ($f_{\rm jet} \sim 10^{-6} - 10^{-4}$) or
the physical conditions necessary to produce bursts are rarely
satisfied in binary mergers and/or the core collapse of massive stars,
implying that the GRB phenomenon is not robust~\cite{lamb99c}.

\section{Burst Energies and Luminosities}

Table 1 lists isotropic-equivalent luminosities and energies of the
bursts for which this information is known.  The maximum energy
$({E_{\rm GRB}})_{\rm max}$ that has been observed for a GRB imposes an
important requirement on GRB models, and is therefore of great
interest to theorists.  The current record holder is GRB 990123 at
$z=1.6$, which implies $E_{\rm GRB} \sim 2 \times 10^{54} f_{\rm jet}$
erg from its gamma-ray fluence, assuming $\Omega_M = 0.3$ and
$\Omega_\Lambda = 0.7$~\cite{kulkarni98}.  Even if GRBs are strongly 
collimated, they are still far and away the brightest electromagnetic 
phenomenon in the Universe, as the following comparison illustrates:

\begin{center}
\begin{itemize}
\leftskip 90pt
\item[$\bullet$] $L_{\makebox[0.28in]{\rm SNe}} \la 10^{44}\ 
{\rm erg}\ {\rm s}^{-1}$
\item[$\bullet$] $L_{\makebox[0.28in]{\rm SGR}} \la 10^{45}\ 
{\rm erg}\ {\rm s}^{-1}$
\item[$\bullet$] $L_{\makebox[0.28in]{\rm AGN}} \la 10^{45}\ 
{\rm erg}\ {\rm s}^{-1}$
\item[$\bullet$] $L_{\makebox[0.28in]{\rm GRB}} \sim  10^{51}\ 
(f_{\rm jet}/10^{-2})\ {\rm erg}\ {\rm s}^{-1}$
\end{itemize}
\end{center}

\begin{table}
\begin{center}
\caption{Peak Photon Fluxes and Isotropic Luminosities for GRBs 
with Known Redshifts}
\begin{tabular}{ccccc} 
\hline
GRB & Redshift & $P$ (ph cm$^{-2}$ s$^{-1}$)$^{a}$ & $L_P$
(ph s$^{-1}$)$^{b}$ & Redshift Reference \\
\hline
970228 & 0.695 & 3.5 & $5.1 \times 10^{57}$ & \cite{djorgovski99b} \\
970508 & 0.835 & 1.2 & $2.5 \times 10^{57}$ & \cite{metzger97} \\
971214 & 3.418 & 2.3 & $6.4 \times 10^{58}$ & \cite{kulkarni98} \\
980613 & 1.096 & 0.63 & $2.3 \times 10^{57}$ & \cite{djorgovski99a} \\
980703 & 0.967 & 2.6 & $7.4 \times 10^{57}$ & \cite{djorgovski98} \\
990123 & 1.600 & 16.4 & $1.2 \times 10^{59}$ & \cite{kulkarni99} \\
990510 & 1.619 & 8.16 & $6.2 \times 10^{58}$ & \cite{vreeswijk99} \\
990712$^{c}$ & 0.430 & -- & -- & \cite{galama99a} \\
\hline
\end{tabular}
\end{center}
$^{a}$ From J. Norris (http://cossc.gsfc.nasa.gov/cossc/batse/counterparts). 
The listed peak photon flux is that in the energy band 50 - 300 keV. \\
$^{b}$Assuming $H_0 = 65$ km s$^{-1}$ Mpc$^{-1}$, 
$\Omega_M = 0.3$, and $\Omega_{\Lambda} = 0.7$. \\
$^{c}$Peak photon flux not yet reported. 
\end{table}

The luminosities of the GRBs listed in Table 1 span a factor of more
than one hundred; i.e., $\Delta L_{\rm GRB}/L_{\rm GRB} \ga 10^2$.  Thus (if
there was previously any doubt), determination of the redshift
distances of a modest number of GRBs has put to rest once and for all
the idea that GRBs are ``standard candles.''  The extensive studies by
Loredo and Wasserman~\cite{loredo98a,loredo98b} and the study by
Schmidt~\cite{schmidt00} show that the luminosity function for GRBs
could well be much broader.

Even taking a luminosity range $\Delta L_{\rm GRB}/L_{\rm GRB} \ga
10^2$ implies that $\Delta F_{\rm GRB}/F_{\rm GRB} \ga 10^4$, given
the range in the distances of the GRBs whose redshifts are currently
known.  This is far broader than the range of GRB peak fluxes in the
BATSE 4B catalog~\cite{paciesas99}, and implies that the flux
distribution of the bursts extends well below the BATSE threshold.

The breadth of the GRB luminosity function and the narrow range in GRB
distances due to cosmology (wasserman92,lamb99b) suggests that a large
part of the differences between apparently bright and apparently dim
bursts (such as time stretching of burst peaks
~\cite{norris94,fenimore95,stern96}, decreased
variability~\cite{stern99}, and spectral softening~\cite{mallozzi96})
are due to {\it intrinsic} differences between {\it intrinsically}
bright and faint bursts, rather than to, e.g., cosmological time
dilation~\cite{lamb99c}.

\section{Burst Models}

The most widely discussed models of the sources of GRBs involve a black
hole and an accretion disk, formed either through the core collapse of
a massive
star~\cite{woosley93,woosley96,pac98,mac99a,mac99b,khokhlov99,wheel00}
or the coalescence of a neutron star -- neutron star (NS-NS) or neutron
star -- black hole (NS-BH) binary~\cite{pac86,narayan92,meszaros93}. 
The former are expected to occur near or in the star-forming regions of
their host galaxies, while most of the latter are expected to occur
outside of the galaxies in which they were born.

The energies listed in Table 1 are difficult to accommodate in NS-NS or
NS-BH binary merger models without invoking strong collimation of the
burst.  Core collapse (so-called ``collapsar'') models can accommodate
such energies more easily, although energies $\ga 10^{54}$ erg 
require collimation, even assuming a high efficiency for the conversion
of gravitational binding energy into gamma-rays~\cite{woosley99}.

Current models of the bursts themselves fall into three general
categories:  Those that invoke a central engine, those that invoke
internal shock waves in a relativistically expanding outflow, and those
that invoke a relativistic external shock wave.  Dermer~\cite{dermer99}
argues that the external shock wave model explains many of the observed
properties of the bursts.  By contrast, Fenimore~\cite{fenimore99} argues
that  several features of GRBs, such as the large gaps seen in burst
time histories, cannot be explained by the external shock wave model,
and that the bursts must therefore be due either to a central engine or
to internal shocks in a relativistically expanding wind.  Either way,
the intensity and spectral variations seen during the burst must
originate at a central engine.  This implies that the lifetime of the
central engine must in many cases be $t_{\rm engine} \ga 100 - 1000$ s,
which poses a severe difficulty for  NS-NS or NS-BH binary merger
models, if such models are invoked to explain the long, softer,
smoother bursts.

\begin{figure}
\psfig{file=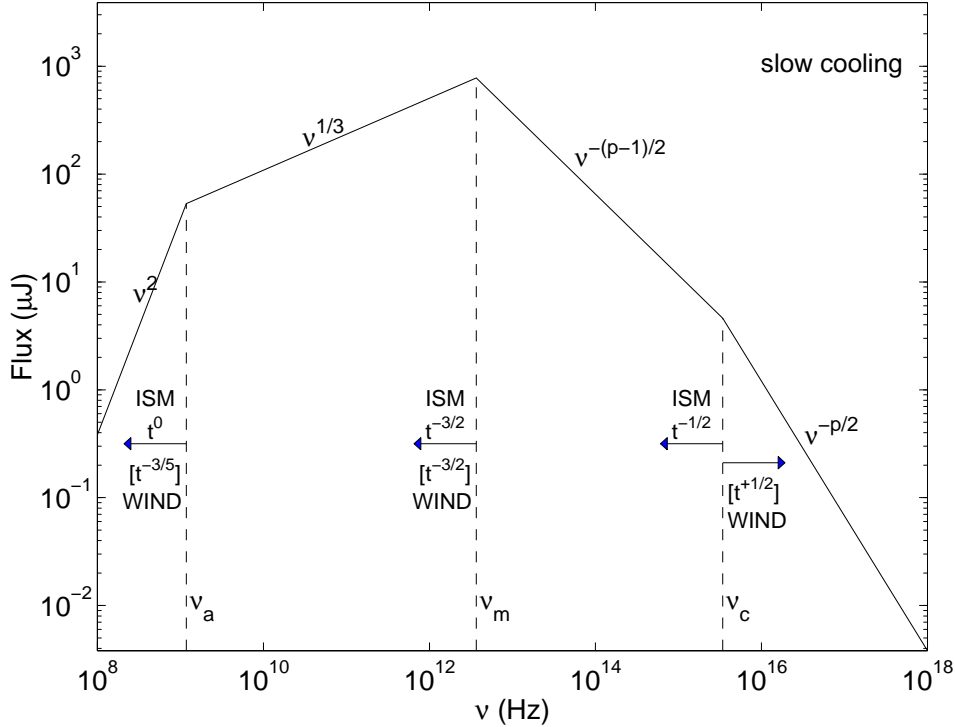,width=5.02truein,clip=}
\caption{Afterglow spectrum expected in the relativistic external shock
model.  The spectrum resembles a broken power-law with four segments
(see text).  The segments are separated by three frequencies:  $\nu_a$,
the synchrotron self-absorption frequency; $\nu_m$, the frequency of
the synchrotron peak, which corresponds to the minimum energy of the
electrons; and $\nu_c$, which corresponds to the energy of the
electrons that have begun to cool significantly.  Note the distinctly
different evolution with time of $\nu_a$ and $\nu_c$ in the ISM and
progenitor wind models.  From~\cite{kulkarni00}.}
\end{figure}

\begin{figure}
\begin{minipage}[t]{2.6truein}
\mbox{}\\
\psfig{file=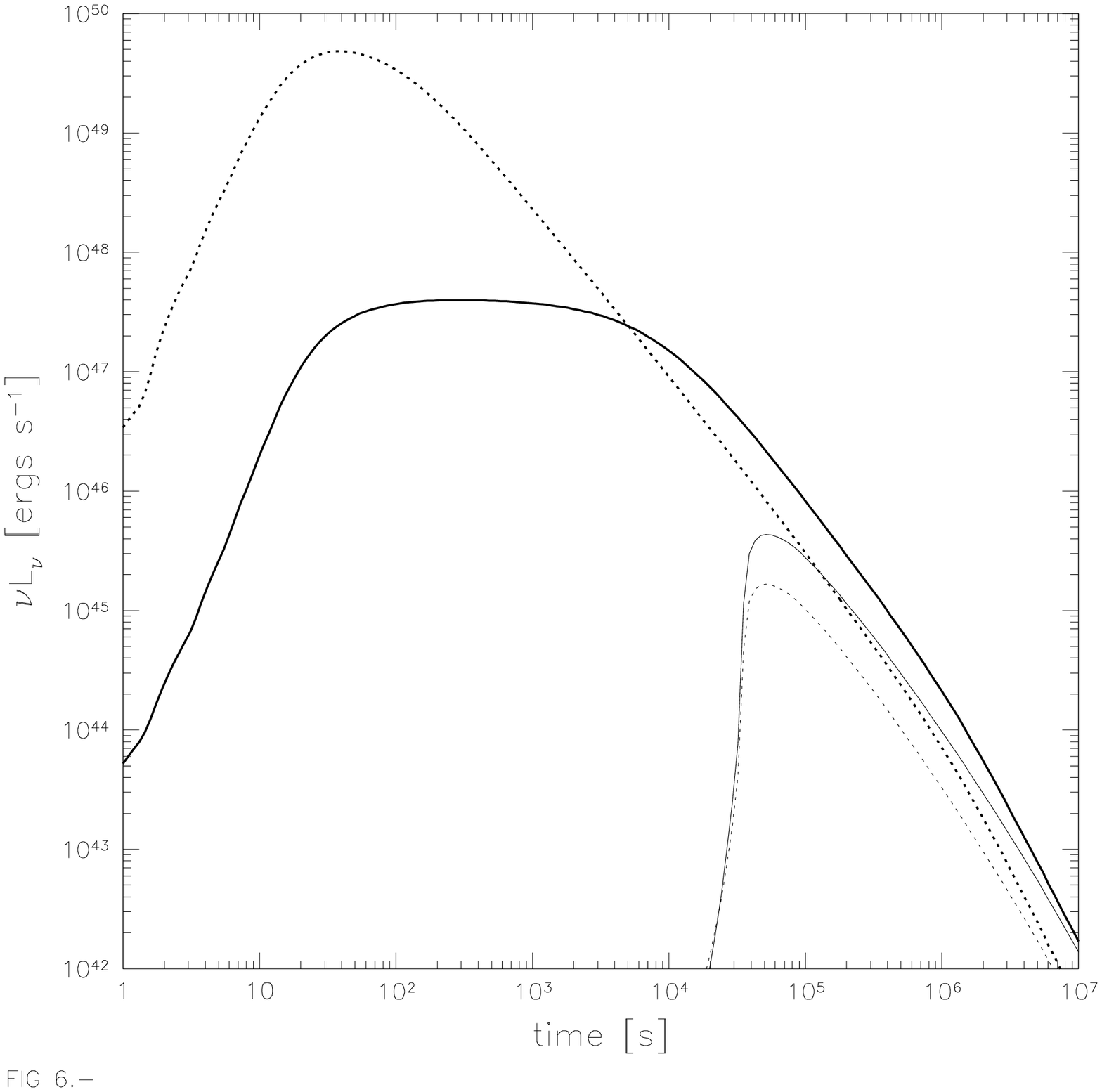,width=2.6truein,clip=}
\caption{Calculations of afterglow light curves for
isotropic-equivalent energy $E_0 = 10^{54}$ ergs s$^{-1}$, $\Gamma_0 =
300$ and $\theta_{jet,0} = 0.2$ radians. The dotted lines are the X-ray
light curves and the solid lines are the optical light curves. The
thick lines are for $\theta_{obs}=0^\circ$ while the thin lines are for
$\theta_{obs}=16^\circ$~\cite{moderski99}.}
\end{minipage}
\hfill
\begin{minipage}[t]{2.6truein}
\mbox{}\\
\psfig{file=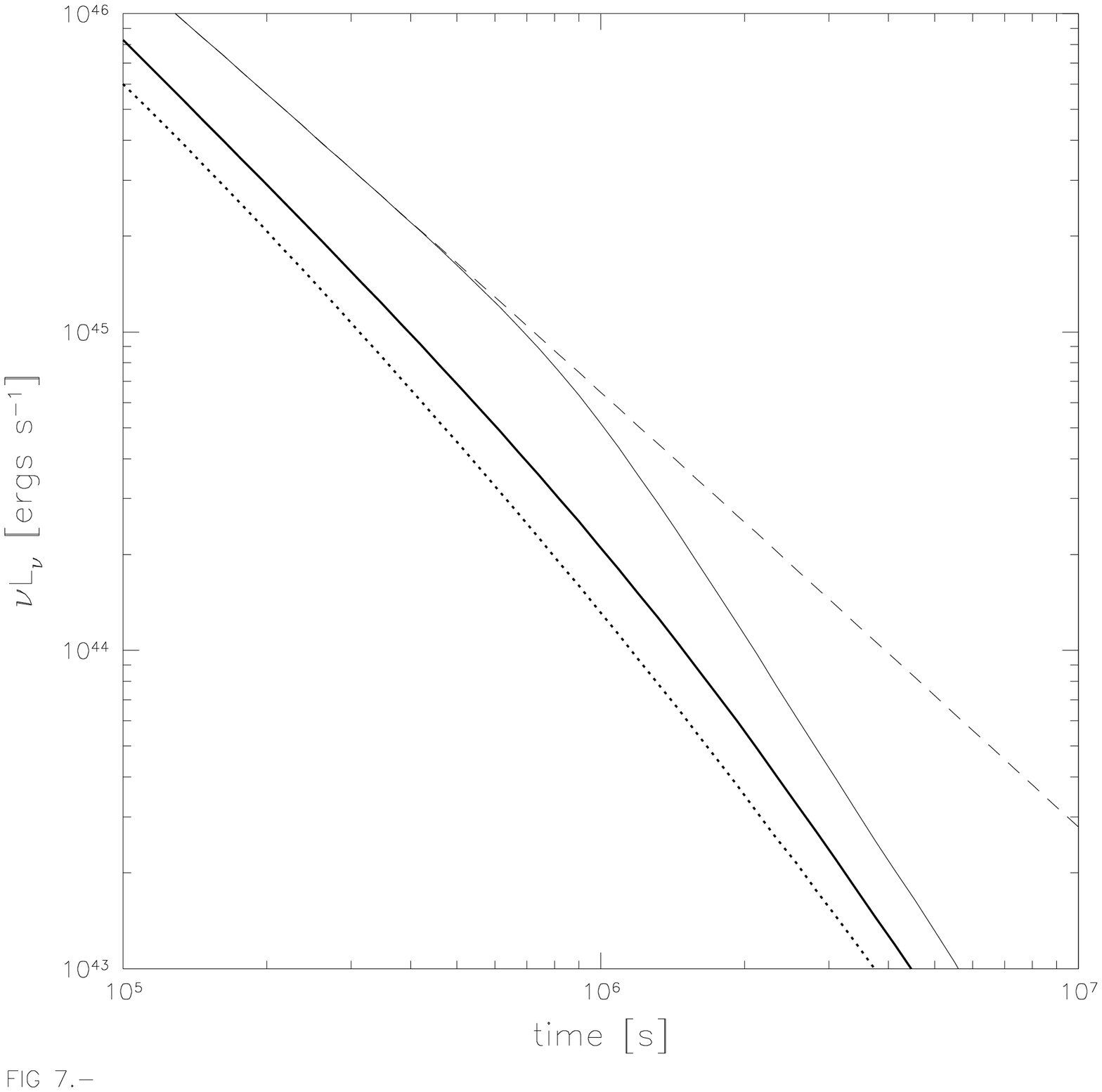,width=2.6truein,clip=}
\caption{Calculations of afterglow light curves in the time range where
the beaming effects are strongest.  The thin solid line is for
$\theta_{\rm jet} = \theta_{\rm jet,0} = 0.2$ radians;  the thick solid
line is for $\theta_{\rm jet} = \theta_{jet,0} + 1/\sqrt 3 \Gamma$; the
dotted line is for $ \theta_{\rm jet} = \theta_{jet,0} + 1/ \Gamma$;
and the dashed line is for $\theta_{\rm jet} = \pi$ (spherical
case)~\cite{moderski99}.}
\end{minipage}
\end{figure}

\section{Afterglow Models}

The most widely discussed model of GRB afterglows is the fireball
model, in which the energy released by the central engine leads to a
relativistic shock wave that expands into the interstellar medium or
into material lost earlier by the burst progenitor via its stellar
wind~\cite{waxman97,sari98}.  As the ejected material encounters the
ambient material, two shocks are produced:  A short-lived reverse shock
that travels through the ejected material and a long-lived forward
shock that propagates into the swept-up ambient material.  In the
fireball model, the  afterglow emission at radio, optical and X-ray
energies is produced by synchrotron emission.  This requires (1)
efficient transfer of energy from the shocked protons (which  contain
most of the energy) to the electrons; (2) efficient acceleration of the
electrons to high energies; and (3) rapid growth of the magnetic field
to values corresponding to energy densities $\sim 10^{-2}$ of that of
the protons.

Figure 10 shows the afterglow spectrum expected in the relativistic
fireball model.  The spectrum resembles a broken power-law with four
segments, which are due to self-absorbed synchrotron emission at radio
energies (segment 1); optically thin synchrotron emission extending from
sub-millimeter through the optical into the UV, with the synchrotron
peak in the near IR or optical (segments 2 and 3); and synchrotron
emission above the synchrotron peak from electrons that have already
cooled (segment 4).  The four broken power-law segments are separated by
three frequencies:  $\nu_a$, the synchrotron self-absorption
frequency; $\nu_m$, the frequency of the synchrotron peak, which
corresponds to the minimum energy of the electrons; and $\nu_c$, which
corresponds to the energy of the electrons that have begun to cool
significantly.

The evolution of the frequencies $\nu_a$ and $\nu_c$ are distinctly
different, depending on whether the relativistic fireball is expanding
into a constant density ISM, or into the decreasing density $\rho \sim
r^{-2}$ expected for material previously ejected by the stellar wind of
the progenitor of the GRB~\cite{chevalier99} (see Figure 10).  

If the relativistic fireball model suffices to explain burst
afterglows, much can be learned (in principle), including the
isotropic-equivalent energy of the fireball, the ratio of the energy
in the magnetic field to that in relativistic electrons, and the
density of the external medium into which the fireball expands (see,
e.g.,~\cite{wijers99,lamb99a}).  This requires, however, enough
simultaneous broad-band photometric or spectral observations to
determine accurately the frequencies $\nu_a,\nu_m$, and $\nu_c$. It
should also be possible, in principle, to use the effects of extinction
due to dust and absorption above the Lyman limit due to hydrogen in the
host galaxy to determine the redshift of the  burst itself, but so far,
this goal has eluded modelers (see, e.g,~\cite{lamb99a}).

\section{Beaming}

Many powerful astrophysical sources produce jets, including young
protostars; so-called ``microquasars,'' which are black hole binaries
in the Galaxy; radio galaxies; and active galactic nuclei.  There is
also evidence from polarization observations that some  supernovae are
asymmetric~\cite{wheel99b}.  Most theorists therefore now expect GRBs
to be significantly collimated, and as discussed above, collimation
reduces the energy required to power the bursts by the collimation
factor $f_{\rm jet}$.  Knowledge of the collimation of GRBs is thus
essential to inferring the energies of the bursts.

In the external shock model of GRB afterglows, the ``edge'' of the jet
will become visible as the shock slows down and relativistic beaming 
decreases.  Furthermore, it is expected that, as the shock slows down,
it will expand laterally, initially at near the speed of
light~\cite{rhoads97}.  The combination of these two effects produces a
steepening of the afterglow light curve (see Figures 11 and 12),
although for the expected values of the parameters involved, the
steepening is rather gradual~\cite{moderski99}.  However, steepening of
the afterglow light curve can also occur as the external shock moves
into denser circumstellar material~\cite{chevalier99,chevalier99b},
confounding the interpretation of the observations.

So far the observational evidence for beaming is inconclusive.  For
example, the lightcurve of the afterglow of GRB 990123 shows a 
steepening after several days that has been interpreted as due to the
spreading of a jet~\cite{meszaros99,kulkarni99,fruchter99b}, but the
break is not achromatic, as would be expected if it were beamed.  On
the other hand, the optical and radio afterglows of GRB 990510 appear
to provide stronger evidence that the burst is
collimated~\cite{kulkarni00}.  However, the absence of large numbers of
``orphan afterglows;'' i.e, X-ray, optical and radio transients
unassociated with GRBs~\cite{greiner99,grindlay99,rhoads97}, argues
that the collimation cannot be extreme.

\begin{figure}
\psfig{file=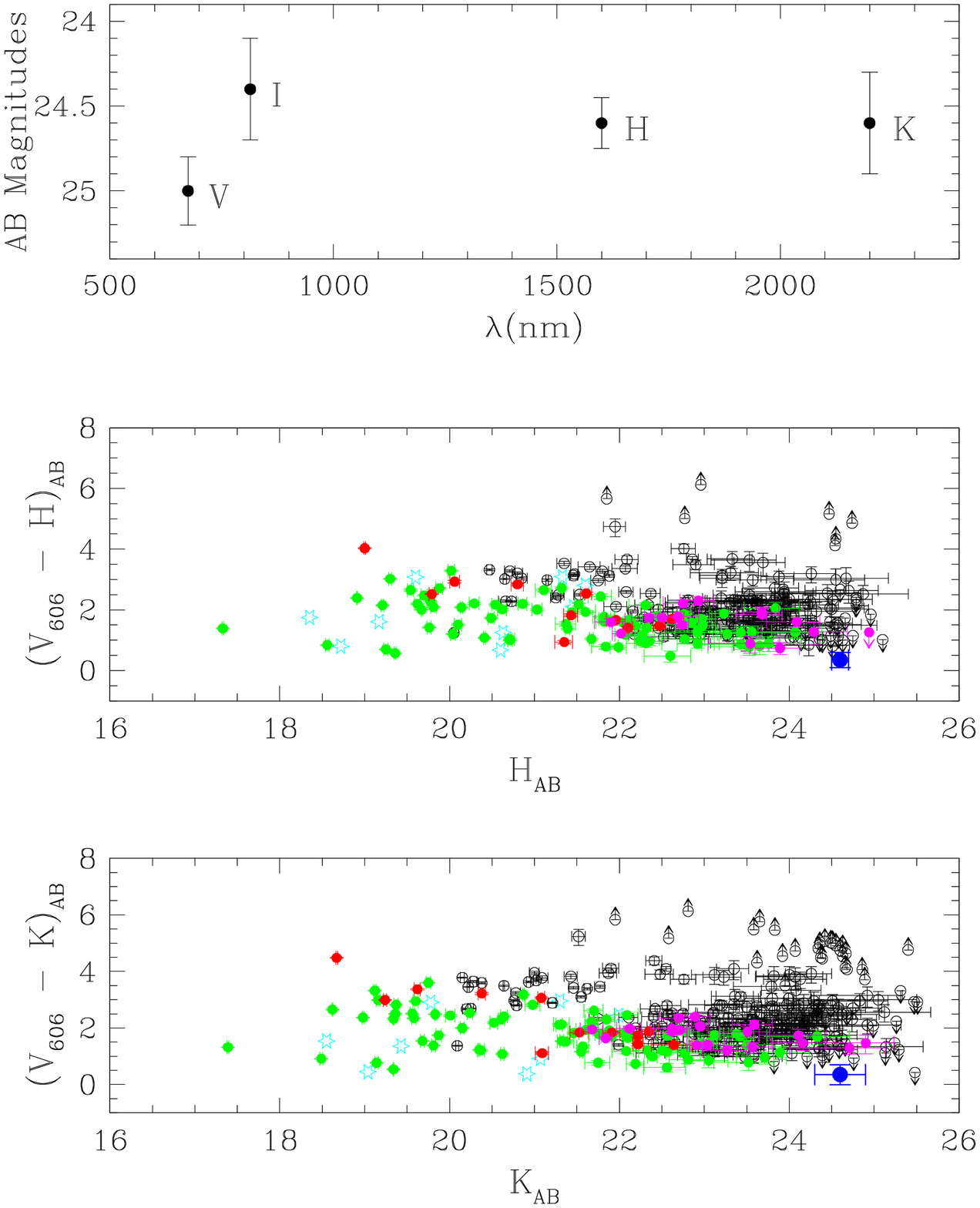,width=5.truein,clip=} 
\caption{HST color of the host galaxy of GRB 970228 (large filled black
circle at the lower right of both panels), compared to the colors of
other galaxies in the GRB field (open circles) and Hubble Deep Field
(filled grey circles).~\cite{fruchter99a}}
\end{figure}

\begin{figure}[b]
\begin{minipage}[t]{2.86truein}
\mbox{}\\
\psfig{file=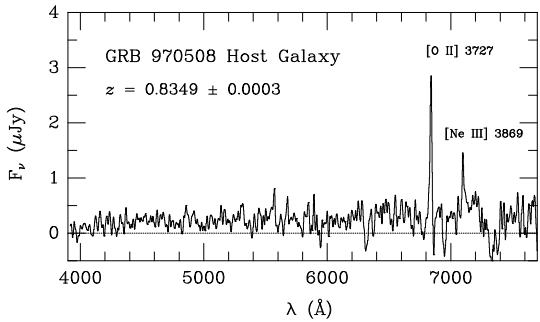,width=2.86truein,clip=}
\end{minipage}
\hfill
\begin{minipage}[t]{2.26truein}
\mbox{}\\ 
\caption{Optical spectrum of the host galaxy of GRB 970508, showing 
[OII] 3727 \AA\ and [Ne III] 3869 \AA\ emission lines redshifted to $z
=0.83$, indicating that the host galaxy is undergoing star formation
and confirming that this is the redshift of the burst~\cite{bloom98}.}
\end{minipage}
\end{figure}

\begin{figure}[t]
\begin{minipage}[b]{2.56truein}
\mbox{}\\
\psfig{file=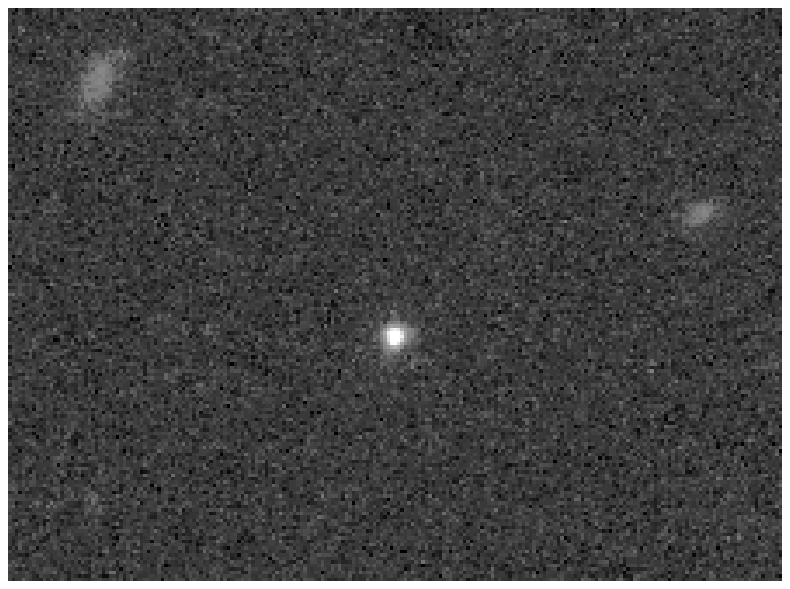,width=2.56truein,clip=}
\end{minipage}
\hfill
\begin{minipage}[b]{2.56truein}
\mbox{}\\
\psfig{file=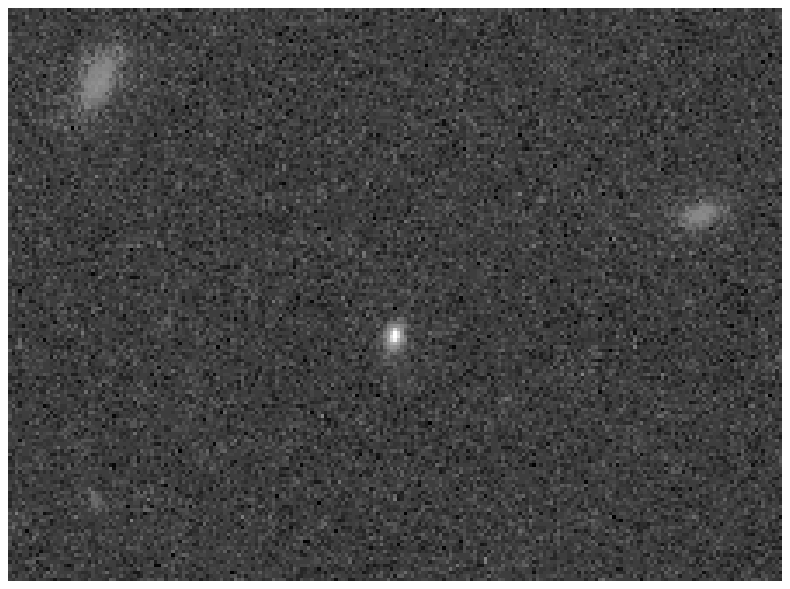,width=2.56truein,clip=}
\end{minipage}
\caption{Left panel: HST STIS image of the host galaxy and afterglow of
GRB 970508 in 1997 June.  Right panel:  HST STIS image of the host
galaxy and afterglow of GRB 970508 in 1997 August showing the continued
fading of the optical afterglow and its coincidence with the region of
star formation in the host galaxy~\cite{pian98}.}  
\end{figure}

\begin{figure}[b]
\begin{minipage}[b]{2.56truein}
\mbox{}\\
\psfig{file=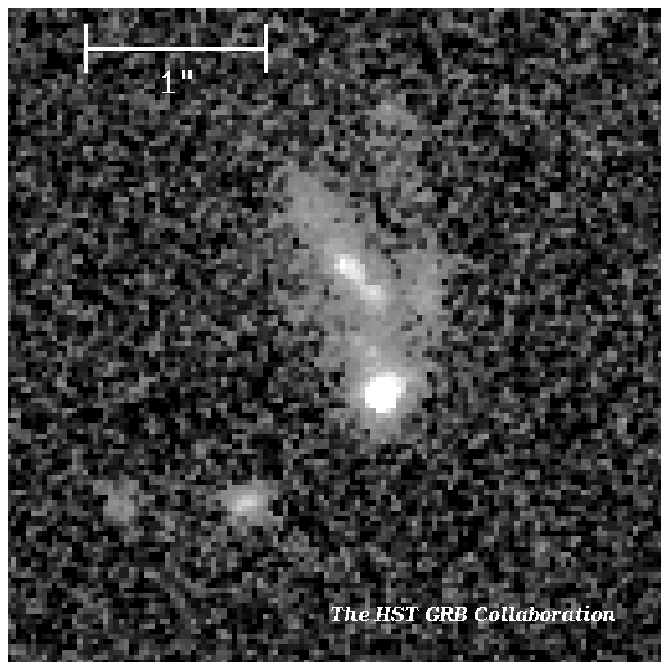,width=2.56truein,clip=}
\end{minipage}
\hfill
\begin{minipage}[b]{2.56truein}
\mbox{}\\
\psfig{file=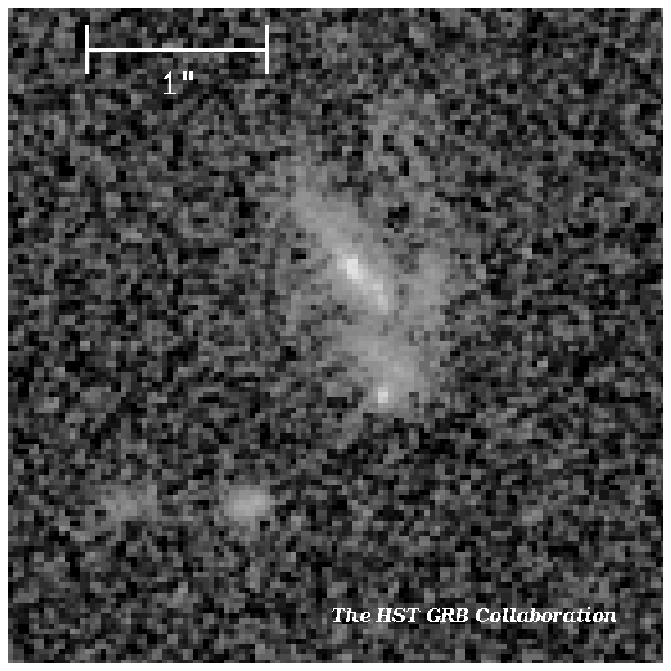,width=2.56truein,clip=}
\end{minipage}
\vfil
\caption{HST STIS observations of the afterglow and host galaxy of GRB
990123, taken in February 1999 (left panel) and in March 1999 (right
panel), showing the continued fading of the optical afterglow and its
coincidence with the region of star formation in the host
galaxy~\cite{fruchter99b}.}
\end{figure}

\section{The GRB-Supernova Connection}

Castander and Lamb~\cite{castander98} showed that the light from the
host galaxy of GRB 970228, the first burst for which an afterglow was
detected, is very blue.  This implies that the host galaxy is
undergoing copious star formation and suggests an association between
GRB sources and star-forming galaxies.  Subsequent analyses of the
color of this  galaxy~\cite{castander99,fruchter99a} (see Figure 13) and
other host galaxies (see, e.g, ~\cite{kulkarni99,fruchter99b} have
strengthened this conclusion, as has the detection of [OII] and
Ly${\alpha}$ emission lines from several host 
galaxies~\cite{metzger97,kulkarni98,bloom98} (see Figure 14).

\begin{figure}
\begin{minipage}[t]{2.56truein}
\mbox{}\\
\psfig{file=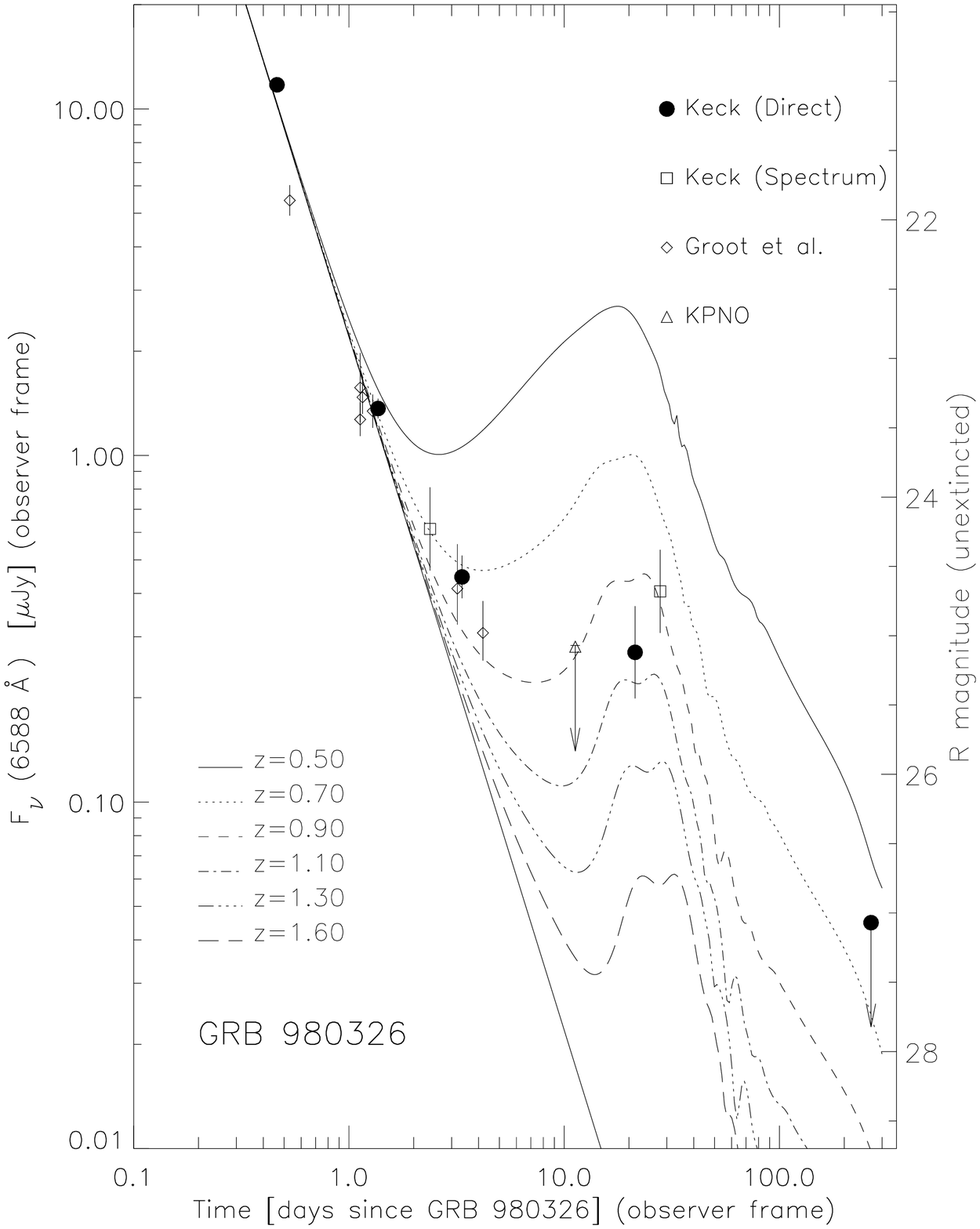,width=2.56truein,clip=}
\end{minipage}
\hfill
\begin{minipage}[t]{2.56truein}
\mbox{}\\ 
\caption{R-band light curve of GRB 980326 and the sum of an
initial power-law decay plus Type Ic supernova light curve for
redshifts ranging from $z = 0.50$ to $z = 1.6$, showing evidence for a 
possible SN component. From~\cite{bloom99}}
\end{minipage}
\end{figure}

The inferred size ($R \la 1 - 3$ kpc) and the morphology of GRB
host galaxies strongly suggest that they are primarily low-mass ($M
\la 0.01 M_{\rm Galaxy}$) but not necessarily sub-luminous
galaxies, because of the ongoing star formation in them (most have $L
\la 0.01 - 0.1 L_{\rm Galaxy}$, but some have $L \sim L_{\rm
Galaxy}$; here $M_{\rm Galaxy}$ and $L_{\rm Galaxy}$ are the mass and
luminosity of a galaxy like the Milky Way).  Thus, it is sometimes not
fully appreciated that, while the total star formation rate in GRB host
galaxies is often modest (resulting in modest [OII] and Ly${\alpha}$
emission line strengths), the star formation rate {\it per unit mass}
in them is very large.

Figures 6, 15 and 16 show HST WFPC2 and STIS images of the optical
afterglows and host galaxies of GRBs 970228, 970508 and 990123.  These
images illustrate the fact that all of the GRB afterglows so far
detected are coincident with bright blue regions of the host
galaxies~\cite{hogg99}.  The positional coincidence between burst
afterglows and the bright blue regions of the host galaxies, and the
evidence for extinction by dust of some burst afterglows (see,
e.g.,~\cite{kulkarni98,lamb99a,reichart98}, suggests that the sources
of these GRBs lie near or in the star-forming regions themselves. 

The increasingly strong evidence that the bursts detected by BeppoSAX
originate in galaxies undergoing star formation, and may occur near or
in the star-forming regions themselves, favors the collapsar model and
disfavors the binary merger model as the explanation for long, softer,
smoother bursts.  Simulations of the kicks given to NS-NS and NS-BH
binaries by the SNe that form them shows that most binary mergers are
expected to occur well outside any galaxy~\cite{bulik99}.  This is
particularly the case, given that the GRB host galaxies identified so
far have small masses, as discussed earlier, and therefore low escape
velocities.  The fact that all of the optical afterglows of the
BeppoSAX bursts are coincident with the disk of the host galaxy
therefore also disfavors the binary merger model as the explanation for
the long, softer, smoother bursts.  However, this evidence is indirect.

\begin{figure}[b]
\hfil
\psfig{file=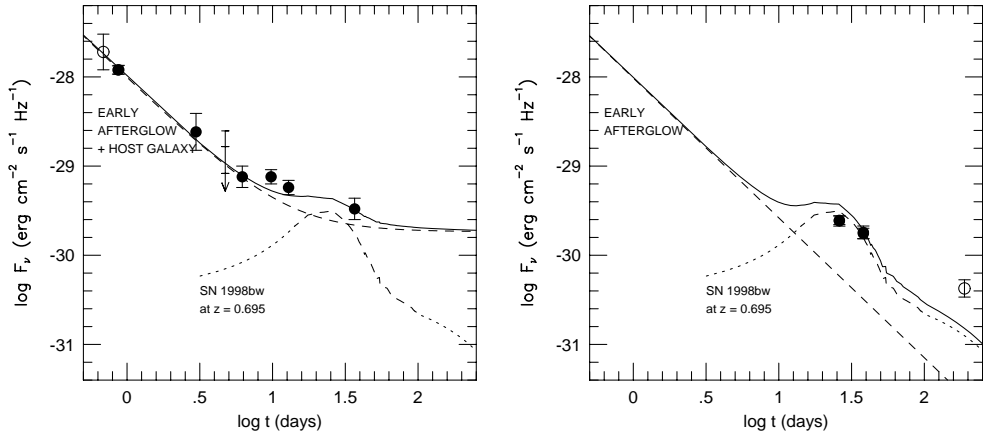,width=5.12truein,clip=}
\caption{Left:  Light curve of GRB 970228 afterglow, showing
ground-based observations.  Right:  Light curve of GRB 970228
afterglow, showing HST observations.  Clear evidence for a second,
possible SN, component in the light curve is seen, particularly in the
HST observations.~\cite{reichart99}}
\hfil
\end{figure}

\begin{figure}[t]
\begin{minipage}[t]{2.56truein}
\mbox{}\\
\psfig{file=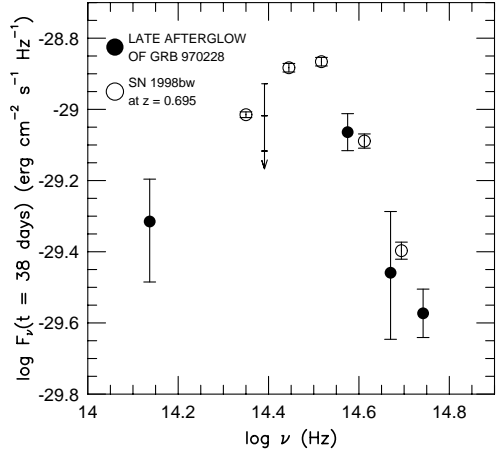,width=2.56truein,clip=}
\end{minipage}
\hfill
\begin{minipage}[t]{2.56truein}
\mbox{}\\
\vspace{-0.5in}
\caption{Broad-band photometric spectrum of possible SN component in
the late afterglow of GRB 970228~\cite{reichart99,reichart00}.  The
filled circles and upper limits are the spectral flux densities ((SFDs)
of the late afterglow after subtracting a model of the SFD of the host
galaxy from Figure 18 and correcting for Galactic extinction; the
unfilled circles are the I- through U-band SFDs of SN 1998bw after
transforming to the redshift ($z = 0.695$) of GRB 970228 and correcting
for Galactic extinction~\cite{reichart99}.}
\end{minipage}
\end{figure}

The discovery that the light curves and spectra of the afterglows of
GRB 980326~\cite{bloom99} (see Figure 17) and GRB
970228~\cite{reichart99} (see Figures 18 and 19) appears to contain a
SN component, in addition to a relativistic shock wave component,
provides a more direct clue that the long, softer, smoother bursts
detected by BeppoSAX are the result of the collapse of massive stars. 
However, the weight of this evidence is limited by the sparseness of
the existing data and the small number of studied afterglows. 

These various discoveries favor the supernova model for the long,
softer, smooth\-er GRBs that comprise 80\% of GRBs and all of the bursts
detected by BeppoSAX to date. However, it is important to remember that
we currently have no clues whatsoever about the nature of the sources
of the short, harder, more variable bursts.  Even their distance scale
is unknown.  Happily, this  situation is likely to change profoundly
with the launch of HETE-2, which is capable of providing accurate,
near-real time positions for both short and long bursts~\cite{lamb00}.

\section{Detectability of GRBs and Their Afterglows at Very High
Redshifts} 

One thing is now clear:  GRBs are a powerful probe of the high-$z$
universe.  Lamb and Reichart~\cite{lamb99b} have calculated the
limiting redshifts detectable by BATSE and HETE-2, and by {\it Swift},
for the seven GRBs with known redshifts and published peak photon
number fluxes (see Table 1).  They find that BATSE and HETE-2 would be
able to detect four of these bursts (GRBs 970228, 970508, 980613, and
980703) out to redshifts $2 \la z \la 4$, and three (GRBs
971214, 990123, and 990510) out to redshifts of $20 \la z \la 30$ (see
Figure 20). {\it Swift} would be able to detect the former four out to
redshifts of $5 \la z \la 15$, and the latter three out to redshifts in
excess of $z \approx 70$, although it is unlikely that GRBs occur at
such extreme redshifts.  Consequently, if GRBs occur at very high
redshifts (VHRs), BATSE has probably already detected them, and future
missions should detect them as well.

\begin{figure}
\begin{minipage}[t]{2.56truein}
\mbox{}\\
\psfig{file=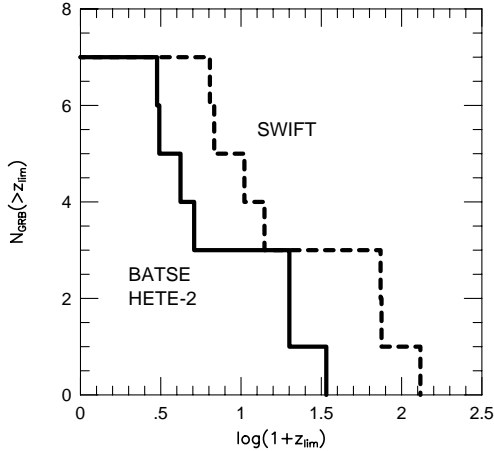,width=2.56truein,clip=}
\end{minipage}
\hfill
\begin{minipage}[t]{2.56truein}
\mbox{}\\
\caption{Cumulative distributions of the limiting redshifts at which
the seven GRBs with well-determined redshifts and published peak photon
number fluxes would be detectable by BATSE and HETE-2, and by {\it
Swift}.~\cite{lamb99b}}
\end{minipage}
\end{figure}

Lamb and Reichart~\cite{lamb99b} also show, somewhat surprisingly, that
the soft X-ray, optical and infrared afterglows of GRBs are detectable
out to VHRs.  The reason is that, while the increase in distance and
the redshifting of the spectrum tend to reduce the spectral flux in GRB
afterglows in a given frequency band, cosmological time dilation tends
to increase it at a fixed time of observation after the GRB, since
afterglow intensities tend to decrease with time.  These effects
combine to produce little or no decrease in the spectral energy flux
$F_{\nu}$ of GRB afterglows at redshifts larger than $z \approx 3$.

This result can be understood as follows.  The spectral flux $F_{\nu}$
of GRB afterglows in a given frequency band and at a fixed time of
observation after the GRB is given as a function of redshift by
\begin{equation}
F_{\nu}(\nu,t) = \frac{L_{\nu}(\nu,t)}{4\pi D^2(z) (1+z)^{1-a+b}},
\end{equation}
where $L_\nu \propto \nu^at^b$ is the intrinsic spectral luminosity of
the GRB afterglow, which we assume applies even at early times, and
$D(z)$ is the comoving distance to the burst.  Many afterglows fade
like $b \approx -4/3$, which implies that $F_{\nu}(\nu,t) \propto
D(z)^{-2} (1+z)^{-5/9}$ in the simplest afterglow model where $a =
2b/3$~\cite{wijers97}.  In addition, $D(z)$ increases very slowly with
redshift beyond a redshift $z \approx 3$.  Consequently, there is
little or no decrease in the spectral flux of GRB afterglows with
increasing redshift beyond $z \approx 3$.  

For example,\ \cite{halpern99} find in the case of GRB 980519 that $a =
-1.05\pm0.10$ and $b = -2.05\pm0.04$ so that $1-a+b = 0.00 \pm 0.11$,
which implies no decrease in the spectral flux with increasing
redshift, except for the effect of $D(z)$.  In the simplest afterglow
model where $a = 2b/3$, if the afterglow declines more rapidly than $b
\approx 1.7$, the spectral flux actually {\it increases} as one moves
the burst to higher redshifts! 

As second example, the left panel of Figure 21 shows the K-band light
curves of the early afterglow of GRB 970228~\cite{reichart99}, as
observed one day after the burst, transformed to various redshifts. 
The transformation involves (1) dimming the afterglow, (2) redshifting
its spectrum, (3) time dilating its light curve, and (4) extinguishing
the spectrum using a model of the Ly$\alpha$ forest.  The left panel of
Figure 21 shows that in this case also, the three effects nearly cancel
one another out at redshifts greater than a few.  Thus the afterglow of
a GRB occurring at a redshift slightly in excess of $z = 10$ would be
detectable at K $\approx 16.2$ mag one hour after the burst, and at K
$\approx 21.6$ mag one day after the burst, if its afterglow were
similar to that of GRB 970228 (a relatively faint afterglow).  

The right panel of Figure 21 shows the resulting spectral flux
distribution.  The spectral flux distribution of the afterglow is cut
off by the Ly$\alpha$ forest at progressively lower frequencies as one
moves out in redshift.  Thus  the redshift distance of high-redshift
($2 \la z \la 5$) GRBs can readily be  determined from the wavelength
of the optical ``dropout'' in their afterglow spectrum and of very high
redshift ($z \ga 5$) GRBs from the wavelength of the infrared
``dropout,'' in their afterglow spectrum~\cite{fruchter99c,lamb99b}. 
Figure 22 shows the visibility of any SN remnant, as a function of
redshift.

These results show that, if GRBs occur at very high redshifts, both
they and their afterglows would be detectable.

\begin{figure}[t]
\begin{minipage}[t]{2.56truein}
\mbox{}\\
\psfig{file=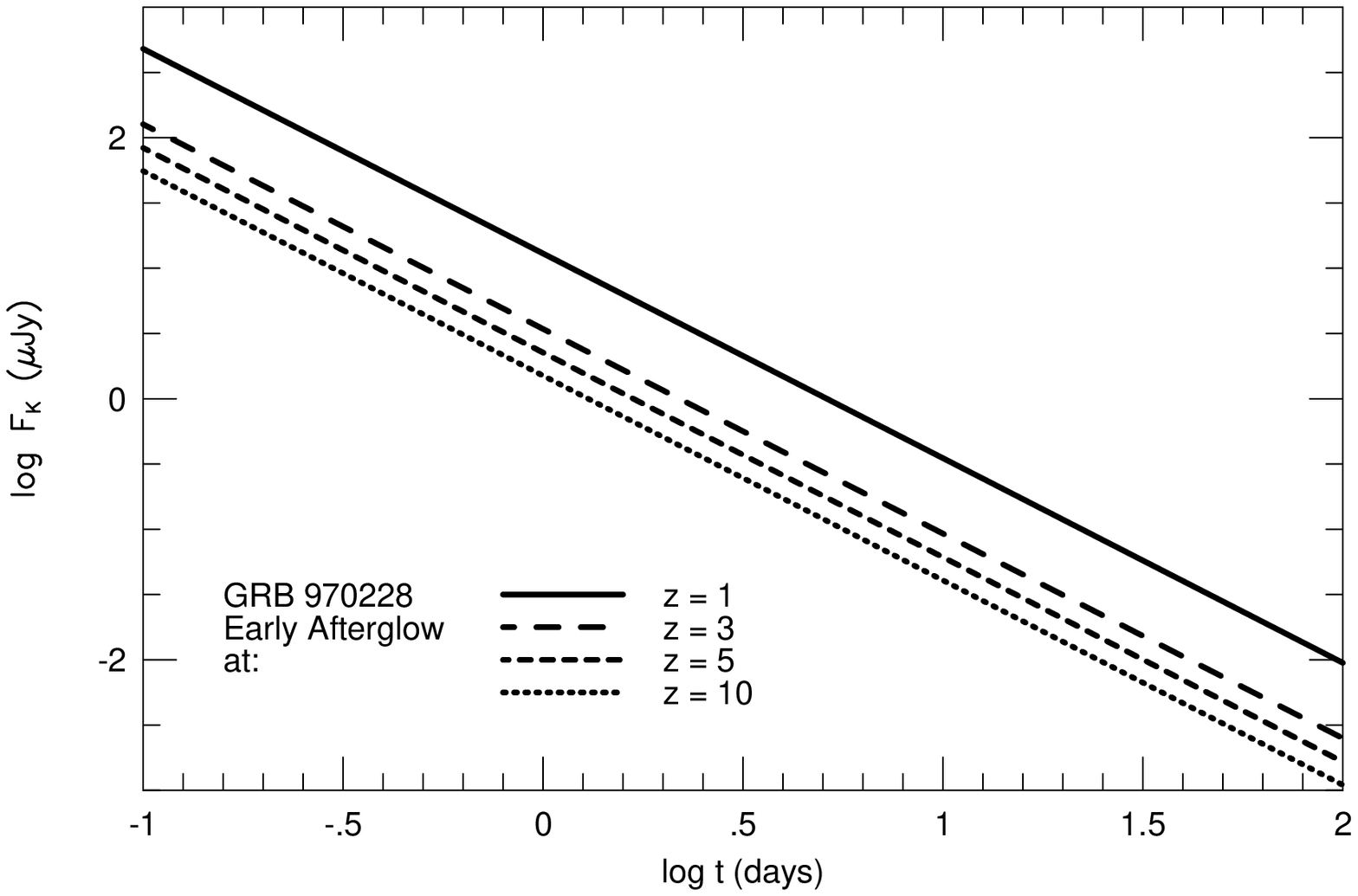,width=2.56truein,clip=}
\end{minipage}
\hfill
\begin{minipage}[t]{2.56truein}
\mbox{}\\
\psfig{file=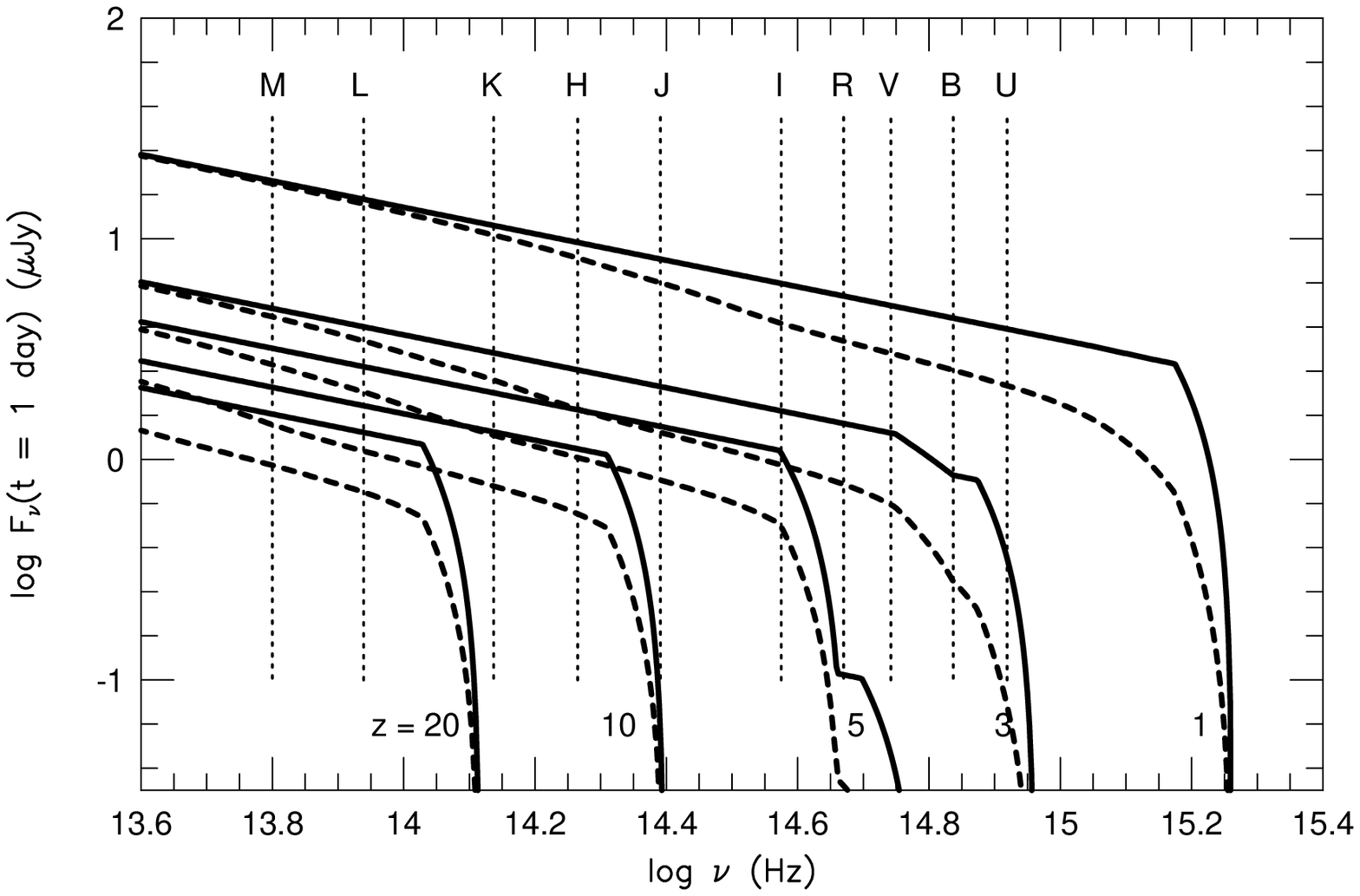,width=2.56truein,clip=}
\end{minipage}
\vfill
\caption{Left panel:  The best-fit light curve of the early afterglow
of GRB 970228, transformed to various redshifts.  Right panel:  The
best-fit spectral flux distribution of the early afterglow of GRB
970228, as observed one day after the burst, after transforming it to
various redshifts, and extinguishing it with a model of the Ly$\alpha$
forest.  The dashed curves are the same as the solid curves, but
extinguished by an amount corresponding to $A_V = 1/3$ mag in the host
galaxy of the burst, using an extinction curve that is typical of the
interstellar medium of our own  galaxy~\cite{lamb99b}.}
\end{figure}

\begin{figure}[b]
\begin{minipage}[t]{2.56truein}
\mbox{}\\
\psfig{file=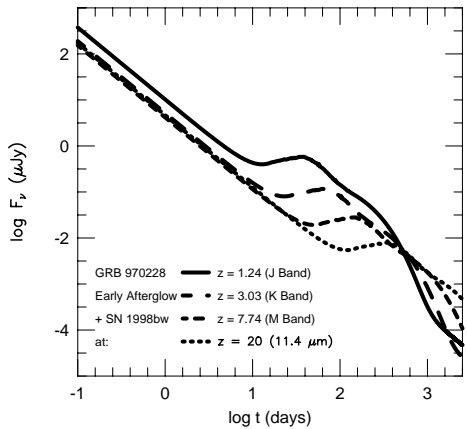,width=2.56truein,clip=}
\end{minipage}
\hfill
\begin{minipage}[t]{2.56truein}
\mbox{}\\
\caption{Light curves of GRB afterglows, with SN components
superimposed, at various redshifts~\cite{lamb99b}.}
\end{minipage}
\end{figure}

\section{GRBs as a Probe of the Very High Redshift Universe}

Observational estimates (see e.g.,~\cite{madau98,r-r99}) suggest that
the star-formation rate (SFR) in the universe was an order of magnitude
larger at a redshift $z \approx 1$ than it is today.  The data at
higher redshifts from the Hubble Deep Field (HDF) in the North suggest
a peak in the SFR at $z \approx 1-2$~\cite{madau98}, but the actual
situation is highly uncertain.  Theoretical calculations show that the
birth rate of Pop III stars produces a peak in the SFR in the universe
at redshifts $16 \la z \la 20$, while the birth rate of Pop II stars
produces a much larger and broader peak at redshifts $2 \la z \la
10$~\cite{ostriker96,gnedin97,valageas99}.  Therefore one expects GRBs
to occur out to at least $z \approx 10$ and possibly $z \approx 15-20$,
redshifts that are far larger than those expected for the most distant
quasars.  Consequently GRBs may be a powerful probe of the
star-formation history of the universe, and particularly of the SFR at
VHRs.  

\begin{figure}
\begin{minipage}[t]{2.56truein}
\mbox{}\\
\psfig{file=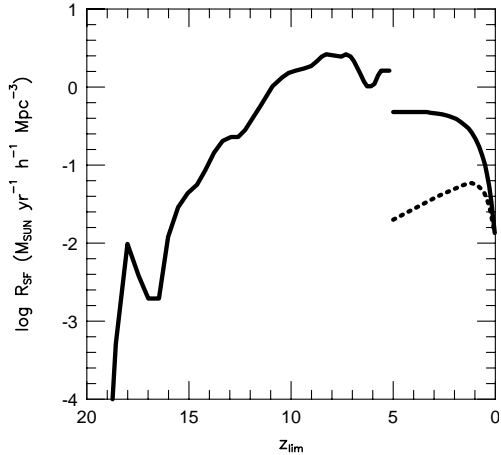,width=2.56truein,clip=}
\end{minipage}
\hfill
\begin{minipage}[t]{2.56truein}
\mbox{}\\
\caption{The cosmic star formation rate (SFR) $R_{SF}$ as a function of
redshift $z$.  The solid curve at $z < 5$ is the SFR derived
by~\cite{r-r99}; the solid curve at $z \ge 5$ is the SFR calculated
by~\cite{gnedin97} (the dip in this curve at $z \approx 6$ is an
artifact of their numerical simulation).  The dotted curve is the SFR
derived by~\cite{madau98}.  From~\cite{lamb99b}.}
\end{minipage}
\end{figure}

Figure 23 shows the SFR versus redshift from a phenomenological
fit~\cite{r-r99} to the SFR derived from sub-millimeter, infrared, and
UV data at redshifts $z < 5$, and from a numerical simulation
by~\cite{gnedin97} at redshifts $z \geq 5$.  The numerical simulations
indicate that the SFR increases with increasing redshift until $z
\approx 8$, at which point it levels off.  The smaller peak in the SFR
at $z \approx 18$ corresponds to the formation of Population III stars,
brought on by cooling by molecular hydrogen.  

Lamb and Reichart~\cite{lamb99b} have calculated the expected number
of GRBs as a function of redshift, assuming (1) that the GRB rate is
proportional to the SFR and (2) that the SFR is that given in Figure
23.  The first assumption may underestimate the GRB rate at VHRs since
it is generally thought that the initial mass function will be tilted
toward a greater fraction of massive stars at VHRs because of less
efficient cooling due to the lower metallicity of the universe at these
early times.  There is a mis-match of about a factor of three between
the $z < 5$ and $z \geq 5$ regimes.  However, estimates of the star
formation rate are uncertain by at least this amount in both regimes. 
Lamb and Reichart~\cite{lamb99b} have therefore chosen to match the two
regimes smoothly to one another, in order to avoid creating a
discontinuity in the GRB peak flux distribution that would be entirely
an artifact of this mis-match.  Figure 24 shows their results.

\begin{figure}[t]
\psfig{file=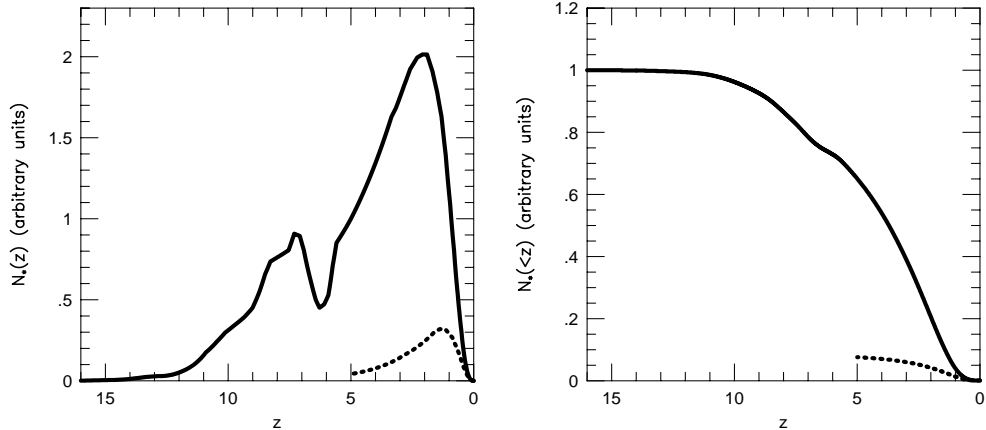,width=5.12truein,clip=} 
\caption{Left panel:  The number $N_*$ of stars expected as a function
of redshift $z$ (i.e., the SFR from Figure 1, weighted
by the differential comoving volume, and time-dilated) assuming that
$\Omega_M = 0.3$ and $\Omega_\Lambda = 0.7$.  Right panel:  The
cumulative distribution of the number $N_*$ of stars expected as a
function of redshift $z$.  Note that $\approx 40\%$ of all stars have
redshifts $z > 5$.  The solid and dashed curves in both panels have the
same meanings as in Figure 1.  From~\cite{lamb99b}}
\end{figure}

The left panel of Figure 24 shows the number $N_*(z)$ of stars expected
as a function of redshift $z$ (i.e., the SFR, weighted by the co-moving
volume, and time-dilated) for an assumed cosmology $\Omega_M = 0.3$ and
$\Omega_\Lambda = 0.7$ (other cosmologies give similar results).  The
solid curve corresponds to the star-formation rate in Figure 23.  The
dashed curve corresponds to the star-formation rate derived
by~\cite{madau98}.  This figure shows that $N_{GRB}(z) [\propto
N_*(z)]$ peaks sharply at $z \approx 2$ and then drops off fairly
rapidly at higher $z$, with a tail that extends out to $z \approx 12$. 
The rapid rise in $N_*(z)$ out to $z \approx 2$ is due to the rapidly
increasing volume of space.  The rapid decline beyond $z \approx 2$ is
due almost completely to the ``edge'' in the spatial distribution
produced by the cosmology.  In essence, the sharp peak in  $N_*(z)$ at
$z \approx 2$ reflects the fact that the SFR is fairly broad in $z$,
and consequently, the behavior of $N_*(z)$ is dominated by the behavior
of the co-moving volume $dV(z)/dz$; i.e., the shape of $N_*(z)$ is due
almost entirely to cosmology.  The right panel in Figure 24 shows the
cumulative distribution $N_*(>z)$ of the number of stars expected as a
function of redshift $z$.  The solid and dashed curves have the same
meaning as in the upper panel.  This figure shows that $\approx 40\%$
of all stars have redshifts $z > 5$.  Since GRBs are detectable at
these VHRs and their redshifts may be measurable from the
absorption-line systems and the Ly$\alpha$ break in the afterglows, if
the GRB rate is proportional to the SFR, then GRBs could provide unique
information about the star-formation history of the VHR
universe~\cite{lamb99b}.

GRB afterglows can also serve as a probe of many other important
aspects of the VHR universe~\cite{lamb99b}.  The absorption-line
systems and the Ly$\alpha$ forest visible in the spectra of GRB
afterglows can be used to trace the evolution of metallicity in the
universe, and to probe the large-scale structure of the universe at
very high redshifts.  Finally, measurement of the Ly$\alpha$ break in
the spectra of GRB afterglows can be used to constrain, or possibly
measure, the epoch at which re-ionization of the universe occurred,
using the Gunn-Peterson test~\cite{gunn65}.

\section{Conclusions}

The discoveries that GRBs have X-ray, optical and radio afterglows have
connected the study of GRBs to the rest of astronomy, and
revolutionized the field.  We now know that the sources of most (and
perhaps all) GRBs lie at cosmological distances, and that the bursts
are among the most energetic and luminous events in the universe. 
There is increasing indirect evidence that the long, softer, smoother
GRBs detected by BeppoSAX are associated with the star-forming regions
of galaxies, and tantalizing direct evidence that these bursts are
connected to core collapse supernovae.  If these GRBs are due to the
collapse of massive stars, they may be a powerful probe of the very
high redshift universe.

\section*{Acknowledgments}

David Schramm was a wonderfully enthusiastic and supportive colleague. 
His scientific interests ranged over much of astrophysics, including
the mysterious phenomenon of gamma-ray bursts.  I am sure that he would
have been excited about recent observational evidence that appears to
link GRBs and one of his favorite subjects, supernovae; and delighted
that GRBs may be a powerful probe of the early universe, a topic to
which he contributed so much.  This work was supported in part by NASA
contract NAGW-4690.

\end{document}